\documentclass[a4paper,fleqn,usenatbib,onecolumn]{mn2e}

\usepackage[T1]{fontenc}
\usepackage{ae,aecompl}

\usepackage{graphicx}	
\usepackage{amsmath}	
\usepackage{amssymb}	
\usepackage{natbib}	

\title[Structures induced by companions in galactic discs]{Structures induced by companions in galactic discs}

\author[P. E. Kyziropoulos et al.]{
P. E. Kyziropoulos,$^{1}$\thanks{E-mail: pkyzirop@ee.duth.gr}
C. Efthymiopoulos$^{2}$\thanks{E-mail: cefthim@academyofathens.gr}
G. A. Gravvanis,$^{1}$\thanks{E-mail: ggravvan@ee.duth.gr}
and P. A. Patsis$^{2}$\thanks{E-mail: patsis@academyofathens.gr}
\\
$^{1}$Department of Electrical and Computer Engineering, School of Engineering,\\
  Democritus University of Thrace, University Campus, Kimmeria, 67100 Xanthi, Greece\\
$^{2}$Research Center for Astronomy, Academy of Athens, Soranou Efessiou 4, 11527 Athens, Greece\\
}

\date{Accepted XXX. Received YYY; in original form ZZZ}

\pubyear{2016}

\begin{document}
\label{firstpage}
\pagerange{\pageref{firstpage}--\pageref{lastpage}}
\maketitle

\begin{abstract}
Using N-body simulations we study the structures induced on a galactic disc by repeated flybys of a companion in decaying eccentric orbit around the disc. Our system is composed by a stellar disc, bulge and live dark matter halo, and we study the system's dynamical response to a sequence of a companion's flybys, when we vary i) the disc's temperature (parameterized by Toomre's Q-parameter) and ii) the companion's mass and initial orbit. We use a new 3D Cartesian grid code: MAIN (Mesh-adaptive Approximate Inverse N-body solver). The main features of MAIN are reviewed, with emphasis on the use of a new Symmetric Factored Approximate Sparse  Inverse (SFASI) matrix in conjunction with the multigrid method that allows the efficient solution of Poisson's equation in three space variables. We find that: i) companions need to be assigned initial masses in a rather narrow window of values in order to produce significant and more long-standing non-axisymmetric structures (bars and spirals) in the main galaxy's disc by the repeated flyby mechanism. ii) a crucial phenomenon is the antagonism between companion-excited and self-excited modes on the disc. Values of $Q >1.5$ are needed in order to allow for the growth of the companion-excited modes to prevail over the the growth of the disc's self-excited modes. iii) We give evidence that the companion-induced spiral structure is best represented by a density wave with pattern speed nearly constant in a region extending from the ILR to a radius close to, but inside, corotation.
\end{abstract}


\section{Introduction}
The grand-design spiral structure generated in disc galaxies by close flybys of their  companion galaxies is a well known phenomenon of galactic astronomy. The pair M51 - NGC 5195 is an archetypical system of this process, that has served also as a benchmarking model for numerical simulations of interacting galaxies, \citep{Howard:1, Salo:3, Salo:1, Salo:2, Dobbs:1, Querejeta:1}; for more general examples of simulations of the influence of satellites on the disk structure see, e.g. \citep{Sundelius:1, Hernquist:2, Donner:1, Huang:1, Oh:1, Kazantzidis:1, Purcell:1, Struck:1, Widrow:1, Oh:2, Donghia:1}. It has been conjectured that the presently observed spiral structure in our own Galaxy might have been dynamically excited by one or more companion flybys, (see e.g. \citet{Purcell:1} for an investigation of the effect of the Sagittarius dwarf galaxy; for older simulations, see e.g. \citet{Ibata:1}). According to the current cosmological paradigm, a significant fraction of disc galaxies with masses $M=10^{10}$ to $10^{12} M_\odot$ should be surrounded by an environment of smaller satellite galaxies, e.g. \citep{Bird:1}, whose exact mass statistics is not yet well understood (see e.g. \citet{Purcell:2} for a review and \citet{Moetazedian:1}). Many of these satellite galaxies might be dark matter sub-halos whose detection could essentially rely on indirect gravitational effects induced upon the galaxy around which they orbit. However, the dynamical features (e.g. strength, pattern speed, longevity, response, etc.) of non-axisymmetric structures like spiral arms induced on a disc by one or more companions is still far from being understood  (\citet{Dobbs:2}, subsection 2.4). 
 
In the present paper we report an ensemble of results collected by a sequence of about 20 different N-body simulations representing disc galaxies influenced by repeated flybys of smaller companion galaxies.  These results emerged out of testing runs of a newly developed N-body code that we call MAIN (Mesh-adaptive Approximate Inverse N-body code). This code is based on a Cartesian grid Poisson equation solver in the usual configuration space. The use of grid codes is rather common in galactic dynamics. Polar 2D grids have been used extensively to test the growth of non-axisymmetric structures in discs (see e.g. \citet{Sellwood:1} and references there in), while 3D spherical grids can be  used to include the effects of 3D features like bars, bulges, or dark matter halos, \citep{Salo:1}. In the case of Cartesian grids, it is customary to solve the Poisson equation in Fourier space, \citep{b:4}, using the `James algorithm' (see e.g. \citet{b:2}, pp. 133) for the production of fixed boundary conditions instead of periodic ones. In MAIN, instead, we return to the basic idea of solving the Poisson equation directly by finite differences in a spatial rectagular grid, using multipole expansions to define the fixed (Dirichlet) conditions on the boundary. Nevertheless, as detailed in section 2 below, MAIN incorporates two key additional features: i) the use of multigrid techniques, and, more importantly, ii) the use of an efficient Symmetric Factored Approximate Sparse Inverse (SFASI) preconditioning scheme for solving the Poisson equation using finite difference method, \citep{Kyziropoulos:1}. This last feature leads to improved convergence behavior and near optimal-complexity, which, along with optimized memory use, allows using MAIN quite efficiently even on single multicore computing workstations (we note however, that parallel implementations of MAIN will be presented in a forthcoming study). Along with the overall simplicity of Cartesian grids, these features render MAIN promising for future developments, e.g. adding gas dynamics or non-gravitational aspects of galaxy evolution. In Section 2, we make a brief presentation of MAIN. 

Section 3 presents the method used to derive the initial conditions of our simulations. Our model for the main galaxy consists of an exponential disc of mass $5\times 10^{10}M_\odot$ and Sersic bulge of $5\times 10^9M_\odot$ embedded in a live, initially spherical and isotropic Dehnen dark halo, whose mass extends to about $2\times 10^{11}M_\odot$ up to $\sim 50$Kpc, and about $10^{12}M_\odot$ at $\sim$200Kpc. The `temperature' (i.e. Q-profile) of the disc is regulated via local relations for the velocity ellipsoid. As in \citet{Salo:1} we prefer to use an algorithm of initial conditions based on a local determination of the velocity ellipsoid at every point of the disc, instead of an algorithm based on global determination of the distribution function (see e.g. \citet{Kuijken:1, McMillan:1}). This allows to fix in detail a Q-profile accross the disc, which we choose to be rising in the central parts of the galaxy while tending to an asymptotically constant value in the outer parts. This method of producing initial conditions does not lead to a perfect collisionless equilibrium in the start. Nevertheless, few timesteps of isolated N-body integration suffice to establish such an equilibrium with satisfactory accuracy for all practical purposes. 

Regarding the companion satellite, we are presently not focusing on its internal structure, which might play a role in the rate of orbital decay of the satellite towards the main galaxy, \citep{Huang:1, Penarrubia:1}. In our computations, we simply consider the whole satellite to be one more Sersic bulge. On the other hand, a factor of key importance is the relative orbit of the satellite with respect to the center of mass of the main galaxy. This is in general a complicated orbit, determined by i) the time-evolving mass distribution within the main galaxy, which affects, in turn both the potential and the rate of tidal dissipation, and ii) the mass of the satellite, which mainly determines the strength of the interaction with the halo via dynamical friction. Our initial shooting of the satellite is based on a Keplerian approximation, i.e., the shooting velocity would lead to a bound Keplerian orbit in a problem of two bodies with masses equal to the mass of the main galaxy and of the satellite respectively. This criterion is rather rough, but proves to be  sufficient for assigning an initial center-of-mass velocity to the satellite. Finally, we consider orbits inclined with respect to the main galaxy's disc, although not as much as in the case of M51, in which the inclination approaches $80^\circ$, \citep{Hernquist:1, Salo:1}. Here, instead, we choose an initial inclination of $30^\circ$, i.e. the lower limit of Holmberg's distribution, \citep{Holmberg:1}. Simple geometrical arguments suggest that the non-axisymmetric tidally induced features should be differentiated according to the angle at which the satellite crosses the galactic plane at perigalacticon. Exploring the role of this parameter alone requires a separate investigation. However, here we fix our initial conditions insisting to have several (typically four or five) such pericentric passages at the edge of the main disc (i.e. at about 5 exponential scale lengths from the disc's center) before the satellite eventually merges with the main galaxy.  

Section 4 describes the main results. We performed simulations with different number of bodies, and/or grid sizes, but here we report only those of $N=10^7$ bodies with an adaptive grid (see section 2). The description in section 4 comprises several kinds of tests, comparison of galaxies with different (outer asymptotic) values of Q with and without companion, growth of self-excited modes over companion-excited modes, amplitude, longevity, and extent of the bar or spiral structure in the main disc, etc. Finally, section 5 contains a summary of conclusions from the present study.

\section{Mesh-Adaptive Approximate Inverse N-body Method (MAIN)}
Some of the most common techniques in the study of galactic systems are: i) Tree-code techniques, (see e.g. \citet{a:2, b:1}), with computational complexity of O($NlogN$), where N is the number of bodies ii) Fast Multipole methods, (see e.g. \citet{a:4}), that can reach up to O($N$) computational complexity though the exact scaling seems implementation dependent as discussed in the relevant literature, (see e.g. \citet{a:12}), and iii) Mesh-type Methods, (see e.g. \citet{b:4}), with computational complexity that varies from O($nlogn$) to near to O($n$) depending on the solution scheme ($n$ is the number of grid points).

Mesh-type methods, also known as `Particle Mesh' methods, \citep{b:4}, solve the Poisson equation instead of applying the Newton's Gravity law on all body pairs. Applying the Finite Difference method to Poisson equation in three space variables, i.e:
\begin{equation} \label{eq:1}
	\bigtriangledown^2 \Phi =4\pi G\rho (r)
\end{equation}
leads to a large order sparse linear system of algebraic equations. Mesh-based methods have been extensively used mainly in conjunction with the Fast Fourier Transform method (FFT), which however has a computational complexity of O($nlogn$). Herewith, we present a new mesh-adaptive approximate inverse N-body  scheme implemented directly in the physical configuration space, where Dirichlet boundary conditions can be used at the boundaries of the computational box. The Multigrid method in conjunction with the so-called Symmetric Factored Approximate Sparse Inverse, `SFASI', matrix as smoother, \citep{Kyziropoulos:1}, is used as a solution scheme having computational complexity close to O($n$).

\begin{figure}
	\begin{center}
	\includegraphics[scale=0.5]{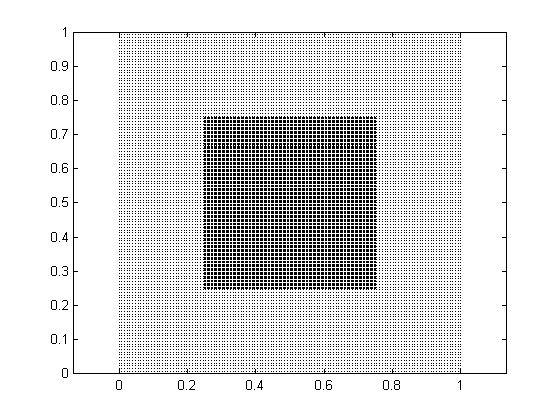}
	\end{center}
	\caption{The adaptive mesh used in the simulations consists of a combination of two overlapping grids with mesh size of 1/256 per dimension from 0 to 1 and 1/512 per dimension from 0.25 to 0.75 respectively.}
	\label{fig:figure1}
\end{figure}

The computational box used for the simulations is the scaled unit box, (1 unit equal to 70 kpc). It should be stated that the quality of the solution in a grid-based N-body method relies on the quality of the grid. Hence, adaptive or semi adaptive schemes have been extensively used in such simulations to avoid excessive computational and memory requirements. In order to simulate the mass-density distribution and solve equation (\ref{eq:1}), an adaptive mesh was used consisting of a combination of two overlapping grids with mesh size of 1/256 per dimension from 0 to 1 and 1/512 per dimension from 0.25 to 0.75 respectively, as depicted in Figure \ref{fig:figure1}. This mesh was formed such that the linear mesh size is set to an approximate value of 136 pc (70kpc/512) on the center of the box where the density of the simulated system is higher.

The right hand side of equation (\ref{eq:1}), which depends on the density, is computed using the Cloud in Cell interpolation method in a three-dimensional grid, \citep{b:4}. The density in the nearest mesh cell from a body ${(x_b, y_b, z_b)}$ with mass ${m_b}$ is computed by
\begin{equation}
	\rho_{i,j,k} =\frac{m_{b}}{h^6}(h-\Delta _{pi})(h-\Delta _{pj})(h-\Delta _{pk})
\end{equation}
where h is the mesh size and ${\Delta _{pi}, \Delta _{pj},\Delta _{pk}}$ is the distance between the body and the center of the mesh cell. Similar formulas hold for the seven remaining mesh cells surrounding the body, \citep{b:4}. On the other hand, to compute the boundary conditions of the above boundary value problem we employ a multipole expansion formula:
\begin{eqnarray} \label{eq:6}
 V &\approx& -\frac{G\sum_{i}^{n}m_i}{R}- \frac{3G}{2R^5}(x^2\sum_{i}^{n}m_ix_i^2+y^2\sum_{i}^{n}m_iy_i^2+z^2\sum_{i}^{n}m_iz_i^2+2xy\sum_{i}^{n}m_ix_iy_i+\nonumber\\
   && +2xz\sum_{i}^{n}m_ix_iz_i+2yz\sum_{i}^{n}m_iy_iz_i)+\frac{G}{2R^3}\sum_{i}^{n}m_i(x_i^2+y_i^2+z_i^2)
\end{eqnarray}
where $G$ is the gravitational constant, $R$ is the distance of the boundary point from the center of mass of the simulated system, $x$, $y$ and $z$ are the coordinates of the boundary point and ${x_i, y_i, z_i}$ and $m_i$ are the coordinates and the mass of the internal nodal points.

The potential at the interior points of the grid is, now, computed by solving the corresponding (sparse) linear system of algebraic equations using the Multigrid method in conjunction with SFASI preconditioning matrix, (more on SFASI in \citet{Kyziropoulos:1}). The Poisson solver implemented in our N-body code (MAIN) is presented in detail in Appendix A.

Once the potential is computed at the grid points, the acceleration in each grid node can be obtained by approximating the derivative of the potential using central differences. Then by re-applying an inverted Cloud in Cell scheme, the acceleration to the bodies is computed by:
\begin{eqnarray}
g_b &=& k_1g_{i,j,k}+k_2g_{i,j,k+1}+k_3g_{i,j+1,k}+k_4g_{i,j+1,k+1}+k_5g_{i+1,j,k}+k_6g_{i+1,j+1,k}+k_7g_{i+1,j,k+1}+k_8g_{i+1,j+1,k+1}
\end{eqnarray}
where the index $b$ denotes the body, g is the accelaration on the mesh cells and ${k_i}$ are computed weights through the Cloud in Cell method, \citep{b:4}. For the nearest mesh cell the $k_1$ is computed by
\begin{equation}
k_1=(h-\Delta_{pi})(h-\Delta_{pj})(h-\Delta_{pk})
\end{equation}
and similar formulas hold for the remaing seven surrounding mesh cells. Hence, the new positions and velocities of the bodies can be computed using the half-step Velocity Verlet integrator, \citep{a:9}. The Velocity Verlet integrator is used widely on N-body simulations due to its low computational complexity, ${O({\Delta t}^2)}$ accuracy and time reversibility.
Considering that the Poisson problem is solved on the unit box, the bodies forced to leave the computational domain must be treated differently. These bodies cannot contribute in the computation of the mass density vector and therefore their potential should be expressed in terms of the overall system using equation (\ref{eq:6}), previously used to compute the boundary conditions. The computation of the potential can be avoided, since the acceleration of these bodies can be directly computed by differentiating equation (\ref{eq:6}), i.e:
\begin{eqnarray}
 \alpha_x &\approx& \frac{-xG\sum_{i}^{n}m_i}{R^3}+\frac{3(3x\sum_{i}^{n}m_ix_i^2+x\sum_{i}^{n}m_iy_i^2+x\sum_{i}^{n} m_iz_i^2)}{2R^5}-\frac{15(x^3\sum_{i}^{n}m_ix_i^2+xy^2\sum_{i}^{n}m_iy_i^2+xz^2\sum_{i}^{n}m_iz_i^2)}{2R^7}- \nonumber\\
    &&  -\frac{15(x^2y\sum_{i}^{n}m_ix_iy_i+x^2z\sum_{i}^{n}m_ix_iz_i+y^2z\sum_{i}^{n}m_iy_iz_i)}{2R^7}+\frac{3(y\sum_{i}^{n}m_ix_iy_i+z\sum_{i}^{n}m_ix_iz_i)}{R^5}
\end{eqnarray}
where $(x,y,z)$ are now the coordinates of the body and $R=(x^2+y^2+z^2)^{1/2}$. The remaining components of the acceleration $a_y$ and $a_z$ can be computed in the same way. 

It should be mentioned that additional functions were implemented for the potential, kinetic energy and angular momentum of the bodies, both inside and outside the computational box, in order to verify the energy and angular momentum conservation of the system, as well as to assess the evolution of the virial ratio. Since the potential inside the computational box is computed by solving the linear system of algebraic equations, the overall potential of the system can be obtained from the values of the mesh grid (since the number of nodes is smaller than the number of bodies, this implies less computational work). Additionally the contribution to the potential energy of the system by particles outside the box is obtained using equation (\ref{eq:6}) on each body multiplied by the mass of the body. Finally, the kinetic energy of the system is computed directly by the sum of $K_b={\frac{1}{2}M_bV_b^2}$ over all bodies.

\section{Initial Conditions}
In our simulations we consider N-body realizations of a `main' and a `companion' galaxy, with initial conditions computed as follows:

\subsection{Main galaxy}
The initial configuration of the main galaxy consists of a dark matter halo, a disc and a bulge. The N-body initial conditions for each component are set as follows:\\
\\
\noindent
{\it Dark matter halo and bulge:} 
We consider an initially spherical dark halo density profile of the form
\begin{equation}\label{eq:halden}
\rho(r) = {\rho_0\over (r/r_h)^{\alpha_h}(1+r/r_h)^{\beta_h-\alpha_h}}  
\end{equation}
with $\alpha_h=1.3$, $\beta_h=3.5$, $r_h=3$Kpc, and 
$$
\rho_0 = {M_h\over 4\pi r_h^3 
\int_0^\infty s^{2-\alpha_h}/(1+s)^{\beta_h-\alpha_h} ds}
$$
where $M_h = 10^{11} M_\odot$. This is a Milky-Way-type halo of medium mass, with a strongly cuspy profile in the center, of inner logarithmic slope $a_h=1.3$; such a power-law is needed in order to correctly capture the influence of the dark halo's inner cusp in structure formation in the disc. On the other hand, the asymptotic outer logarithmic slope is $\beta_h-\alpha_h=2.2$, implying that the mass increases with distance from the center asymptotically (for $r>>r_h$) as $\sim (r/r_h)^{0.8}$. Thus, the halo encloses about $5\times 10^{10}$ $M_\odot$ at $r=10$Kpc, which rises to about $3\times 10^{11}M_\odot$ at $r=100$Kpc. We use Eddington's formula
\begin{equation}\label{eq:eddi}
f(E) = {1\over\sqrt{8}\pi^2} {d\over dE}\int_E^0 {d\rho\over dV} {dV\over\sqrt{V-E}} 
\end{equation}
to compute an isotropic distribution function $f(E)$ corresponding to the density profile (\ref{eq:halden}). In Eq.(\ref{eq:eddi}) we need to know $\rho$ as a function of the gravitational potential $V$. To this end, we first compute a list of values $V_i = V(r_i)$, with $r_i=(0.01\times i)$Kpc, $i=0,\ldots,2500$, where
\begin{equation}\label{eq:halpot}
V(r) = -{G M(r)\over r^2} -G\int_r^\infty {M(s) ds\over s^2}
\end{equation}
with $G = 4.3\times 10^6{Km}^2s^{-1}$Kpc $M_\odot^{-1}$, $M(r) = \int_0^r 4\pi s^2\rho(s)ds$ with $\rho(s)$ substituted by $\rho_h(s)$. Having now the paired list $(\rho_i\equiv\rho_h(r_i),V_i)$, which is equispaced in $r_i$, we produce, via linear interpolation, a corresponding paired list $(\rho_j,V_j)$, but now equispaced in the $V_j$, $j=0,\ldots,2500$. Finally, we compute $(d\rho/dV)_i\approx (\rho_{i+1}-\rho_i)/(V_{i+1}-V_i)$. This allows to obtain also, via (\ref{eq:eddi}), the distribution function as a list of values $f_i\equiv f(E_i)$, where $E_i=(V_i+V_{i+1})/2$. This yields $f(E)\equiv f(E(\mathbf{r},\mathbf{v}))$, and thus, positions and velocities corresponding to the above distribution function can be assigned to the particles via a rejection algorithm, \citep{b:6}. We use $4.5\times 10^6$ particles to simulate the dark matter halo, and we truncate this halo at 50Kpc. 

\begin{figure}
	\begin{center}
	\includegraphics[scale=0.7]{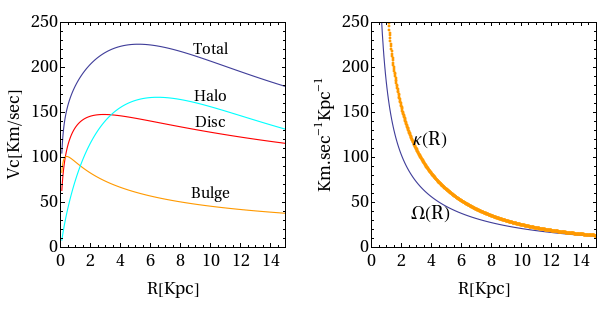}
	\end{center}
	\caption{Left panel: the rotation curve (top curve, blue) resulting from the three components of the main galaxy's initial conditions. The contribution of each component is also shown. Right panel: the angular velocity of the circular orbit $\Omega(R)$ and the epicyclic frequency $\kappa(R)$ for the adopted model's parameters (see text).}
	\label{fig:rotcurve}
\end{figure}

A similar procedure is used in order to produce initial conditions for the bulge particles. Here, we only change the density profile to a Sersic law:
\begin{equation}\label{eq:bulgeden}
\rho_b(r) = -{1\over\pi}\int_r^\infty 
{d\Sigma_b(R)\over dR}{1\over\sqrt{R^2-r^2}} dR
\end{equation}
where
$$
\Sigma_b(R) = \Sigma_0
\exp\left(-b_b\left({R\over r_b}\right)^{1\over n_b}-1\right)
$$
with $b_b=2n_b-0.324$, $n_b=3.5$, and 
$$
\Sigma_0 = {M_b\over \int_0^\infty 2\pi s \exp\left(-b_b\left({s\over r_b}\right)^{1\over n_b}-1\right) ds}
$$
with $M_b = 5\times 10^9 M_\odot$. Here, for the numerical inversion of Eddington's formula we use a grid of points $r_i=(0.01\times i)$Kpc, with $i=0,\ldots,1000)$. Finally, we materialize the bulge with $5\times 10^5$ particles.\\
\\
\noindent
{\it Disc:} We consider an exponential disc with density (in cylindrical co-ordinates $(R,\phi,z))$:
\begin{equation}\label{eq:discden}
\rho_d(R,\phi,z)=\left({M_d\over 8\pi^2 z_dR_d^2}\right)
\exp\left(-{|z|\over z_d}-{R\over R_d}\right) 
\end{equation}
where $M_d = 5\times 10^{10}M_\odot$, $R_d = 3$Kpc, $z_d=0.2$Kpc. We then consider the distribution function
\begin{equation}\label{eq:discdf}
f_d(R,\phi,z,v_R,v_\phi,v_z) = (2\pi)^{-3/2}
\left(\sigma_R(R)\sigma_\phi(R)\sigma_z(R)\right)^{-1/2}\rho_d(R,\phi,z)
\exp\left(-{v_R^2\over 2(\sigma_R(R))^2}
-{(v_\phi-v_{0,\phi})^2)\over 2(\sigma_\phi(R))^2} 
-{v_z^2\over 2(\sigma_z(R))^2}\right)
\end{equation}
where
$$
\sigma_R(R) = {3.36 \Sigma(R)Q(R)\over\kappa(R)},~~~
\sigma_\phi(R)={\kappa(R)\sigma_R(R)\over 2\Omega(R)},~~~
\sigma_z(R)={\kappa(R)\sigma_R(R)\over \kappa_z(R)},
$$
$$
v_{0,\phi}=\left(v_c^2(R)-\sigma_\phi^2(R)
+\sigma_R^2(R) + {d\ln\Sigma(R)\over d\ln R}\right)^{1/2}
$$
(the so-called `asymmetric drift', where $\Sigma(R)=(M_d/(2\pi R_d^2))\exp(-R/R_d)$), and
\begin{equation}\label{eq:qprofile}
Q(R) = Q_\infty+(Q_0-Q_\infty)\exp(-R/Rq) ~~~.
\end{equation}
This last equation fixes the profile of Toomre's Q-parameter to start from an inner value $Q_0$ and fall asymptotically to $Q_\infty$ with an exponential law fixed by the scale parameter $R_q$. We set $Q_0=5.5$, $R_q=2$Kpc, and vary $Q_\infty$ to obtain experiments of discs of various temperatures, i.e., $Q_\infty=0.5,1,1.5,2,2.5$. In the sequel, when referring to a Q-value for a particular experiment we mean the asymptotic outer value $Q_\infty$. Finally, in the above equations, the quantities $\kappa(R)$, $\kappa_z(R)$, $\Omega(R)$ and $v_c(R)$ represent the epicyclic, vertical, rotational frequencies, and velocity of the circular orbit, respectively, under the {\it full} potential on the disc plane. We have
\begin{eqnarray}\label{eq:epic}
&~&\Omega(R)=\left({1\over R}{dV_{tot}(R,z=0)\over dR}\right)^{1/2},~~~
\kappa(R)=\left({d^2V_{tot}(R,z=0)\over dR^2} + 3\Omega^2(R)\right)^{1/2},~~~\\
&~&\kappa_z(R)=\left({d^2V_{tot}(R,z=0)\over dz^2}\right)^{1/2},~~~
v_c(R)=\Omega(R)R \nonumber
\end{eqnarray}
and 
$V_{tot} = V_h+V_b+V_d$, i.e., the sum of the potentials of all three components of the main galaxy. Figure \ref{fig:rotcurve} shows the rotation curve $v_c(R)$ obtained by the total potential as well as the contributions to it of each of the three galaxy components separately. We also show the functions $\kappa(R)$ and $\Omega(R)$ for reference in the sequel.

Using the rejection algorithm implemented on the distribution function (\ref{eq:discdf}), we assign positions and velocities to the disc particles. The disc is materialized using $5\times 10^6$ particles. \\
\\
{\it Virialization:} The two initially spherical components (bulge + halo) are, each, independently, in equilibrium state while the disc distribution function (Eq. (\ref{eq:discdf})) is also approximately so within the total potential. However, when all three components are added in the same system, the total system is not in equilibrium, although neither very far from it. After correcting all particles' positions and velocities for statistical fluctuations, so that the common center of mass is brought at $x_{cm}=y_{cm}=z_{cm}=$ $v_{x,{cm}}=v_{y,{cm}}=v_{z,{cm}}=0$, we find an initial virial ratio $|W|/K \approx 2.8$, where $W$, $K$ are the potential and kinetic energy, respectively, of the total system. In order to avoid exciting instabilities in the disc due to this initial state, we apply a two-stage relaxation process. First, we allow the halo to evolve self-consistently using the N-body code, in which the total potential is computed in every step keeping the positions and velocities of the disc and bulge particles frozen. The halo undergoes a rapid relaxation, which slightly changes the form of its distribution function, deepening the central potential well by a few percent. After a time of the order of $\sim 3\times 10^{-1}Gyr$ the virial ratio of the total system goes very close to 2. Finally, we switch on the time evolution of the bulge+disc particles, and allow them also to adapt their distribution functions to the new halo configuration. This second relaxation lasts for about $10^{-1}Gyr$. We store the final snapshot, which is then the starting point for simulations including the companion galaxy. \\
\\
\noindent
{\it Companion galaxy:} we initially materialize a satellite galaxy by $5\times 10^5$ particles of total mass $M_s$ taking different values between $5\times10^9M_\odot$ and $2\times 10^{10}M_\odot$ (see section 4). Our satellite galaxies are produced in the same way as Sersic bulges, adapting the scale length $r_b$ to larger values for larger masses. Thus, we set $r_b=1.5$Kpc when $M_s=5\times10^9M_\odot$ and $r_b=2.0$Kpc when $M_s=10^{10}M_\odot$ or $M_s=2\times10^{10}M_\odot$. We finally make the center-of-mass correction for the positions and velocities of all the satellite's particles. 

So far, both the main galaxy and the satellite are centered around the origin of the phase space. We now place the satellite at a certain distance from the main galaxy, and assign systematic initial velocities to both sub-systems so that their mutual orbit around the common barycenter resembles, initially, to a Keplerian ellipse of a certain chosen value of the major semi-axis $a$, eccentricity $e$ and inclination $I$. To this end, we first shift the positions of all the particles of the satellite according to the rule:
\begin{equation}\label{eq:satposkep}
x\rightarrow x + a(1+e)\cos I,~~~~y\rightarrow y,~~~~z\rightarrow z+ a(1+e)\sin I 
\end{equation}
Eq.(\ref{eq:satposkep}) places the satellite initially at apocenter of the above `nominal' Keplerian orbit with elements $(a,e,I)$, provided that the satellite is shooted in the y-direction with an appropriate relative velocity $\Delta V$ with respect to the main galaxy. In order, now, to estimate this velocity, let $M_0$ be an estimate of the total mass inside a sphere or radius $a(1+e)$ centered around the main galaxy. We perform experiments with $e=0.4$ and $a=27$ or 
$a=20$Kpc. Taking into account the mass distribution of all the components in the main galaxy, we have that the proportion of dark to luminous matter in our model is somewhat higher than 1:1 at distances up to $~30$Kpc. We then set $M_0 = 1.3\times 10^{11}$, which is somewhat higher than twice the `luminous' mass $M_d$ and $M_b$. Now, for two point masses $M_0~+~M_s$ placed at the origin and at the position of Eq.(\ref{eq:satposkep}) respectively, the relative Keplerian ellipse of elements $(a,e,I)$ is realized if we set 
\begin{equation}\label{dv}
\Delta V = {\sqrt{G(M_0+M_s)a(1-e^2)}\over a(1+e)}
\end{equation}
This velocity cannot be all assigned to the satellite, because then the total system acquires a non-zero center-of-mass velocity in the y-direction. Instead, we change the velocities of all the particles in both the main galaxy and the satellite according to:
\begin{eqnarray}\label{eq:satvelkep}
~&\mbox{Main galaxy: }&v_x\rightarrow v_x,~~~ v_y\rightarrow v_y-{M_s\over M_{tot}}\Delta V,~~~ z\rightarrow z \\
~&\mbox{Satellite: }&v_x\rightarrow v_x,~~~ v_y\rightarrow v_y+{M_s\over M_{tot}}\Delta V,~~~ z\rightarrow z \\
\end{eqnarray}
where $M_{tot}$ is the total {\it true} mass of all the particles (main galaxy + satellite) which are within the computational box. Eq.(\ref{eq:satvelkep}) ensures now that the common barycenter of the system has no velocity with respect to the center of the computational box. 

\begin{table}
\centering
\caption{Main Galaxy Initial Parameters}
\label{table:1}
\begin{tabular}{|ll|ll|ll|}
\hline
\multicolumn{2}{|l|}{Halo}       & \multicolumn{2}{l|}{Bulge} & \multicolumn{2}{l|}{Disc}   \\ \hline
$r_h$ (Kpc)       & 3.0                   & $n_b$              & 3.5            & $z_d$ (Kpc)       & 0.2              \\
$a_h$             & 1.3                   & $r_b$ (Kpc)        & 1.5            & $R_d$ (Kpc)       & 3.0              \\
$b_h$             & 3.5                   & $M_b$ ($M_\odot$)  & $5\times10^9$  & $M_d$ ($M_\odot$) & $5\times10^{10}$ \\
$M_h$ ($M_\odot$) & $10^{11}$ &                    &                &                   &                  \\ \hline
\end{tabular}
\end{table}

\begin{table}
\centering
\caption{List of all experiments. Class A experiments simulate only the main galaxy and class B experiments simulate the main galaxy  interacting with a companion galaxy.}
\label{table:2}
\begin{tabular}{cccccccccccccc}
\multicolumn{1}{c|}{Experiment} & \multicolumn{3}{c|}{Disc's Q profile}     & \multicolumn{3}{c|}{Companion}  & \multicolumn{3}{c|}{Comp. init. orbit} & \multicolumn{4}{c}{$G\times$particle mass ($G M_\odot$)} \\ \hline
No.                             & $R_Q$ (Kpc) & $Q_0$ & $Q_{\infty}$ & $R_c$ (Kpc) & $n_c$ & $M_c$ ($M_\odot$) & a (Kpc)    & e      & I($^\circ$)    & Bulge     & Halo     & Disc     & Comp.   \\
Exp. $A_1$                      & 2.0         & 5.5   & 0.5          & -           & -     & -                 & -          & -      & -    & 0.0412    & 0.0730    & 0.0375    & -         \\
Exp. $A_2$                      & 2.0         & 5.5   & 1.0          & -           & -     & -                 & -          & -      & -    & 0.0412    & 0.0730    & 0.0375    & -         \\
Exp. $A_3$                      & 2.0         & 5.5   & 1.5          & -           & -     & -                 & -          & -      & -    & 0.0412    & 0.0730    & 0.0375    & -         \\
Exp. $A_4$                      & 2.0         & 5.5   & 2.0          & -           & -     & -                 & -          & -      & -    & 0.0412    & 0.0730    & 0.0375    & -         \\
Exp. $A_5$                      & 2.0         & 5.5   & 2.5          & -           & -     & -                 & -          & -      & -    & 0.0412    & 0.0730    & 0.0375    & -         \\
Exp. $B_1$                      & 2.0         & 5.5   & 1.5          & 0.7         & 3.5   & $10^9$            & 27         & 0.4    & 30   & 0.0412    & 0.0730    & 0.0375    & 0.0428    \\
Exp. $B_2$                      & 2.0         & 5.5   & 2.0          & 0.7         & 3.5   & $10^9$            & 27         & 0.4    & 30   & 0.0412    & 0.0730    & 0.0375    & 0.0428    \\
Exp. $B_3$                      & 2.0         & 5.5   & 1.5          & 1.5         & 3.5   & $5\times10^9$     & 27         & 0.4    & 30   & 0.0412    & 0.0730    & 0.0375    & 0.2060    \\
Exp. $B_4$                      & 2.0         & 5.5   & 1.5          & 2.0         & 3.5   & $10^{10}$         & 27         & 0.4    & 30   & 0.0412    & 0.0730    & 0.0375    & 0.3989    \\
Exp. $B_5$                      & 2.0         & 5.5   & 1.5          & 2.0         & 3.5   & $2\times10^{10}$  & 27         & 0.4    & 30   & 0.0412    & 0.0730    & 0.0375    & 0.7978    \\
Exp. $B_6$                      & 2.0         & 5.5   & 2.0          & 2.0         & 3.5   & $10^{10}$         & 27         & 0.4    & 30   & 0.0412    & 0.0730    & 0.0375    & 0.3989    \\
Exp. $B_7$                      & 2.0         & 5.5   & 2.0          & 2.0         & 3.5   & $10^{10}$         & 20         & 0.4    & 30   & 0.0412    & 0.0730    & 0.0375    & 0.3989    \\
Exp. $B_8$                      & 2.0         & 5.5   & 2.5          & 2.0         & 3.5   & $10^{10}$         & 20         & 0.4    & 30   & 0.0412    & 0.0730    & 0.0375    & 0.3989    \\
Exp. $B_9$                      & 2.0         & 5.5   & 2.0          & 2.0         & 3.5   & $2\times10^{10}$  & 20         & 0.4    & 30   & 0.0412    & 0.0730    & 0.0375    & 0.7978   
\end{tabular}
\end{table}

Table \ref{table:1} summarizes those parameters in the initial conditions which remain unaltered in all our simulations. These are the parameters determining the profiles of the main galaxy's bulge, disc, and dark matter halo. Table \ref{table:2} shows, instead, parameters which are varied in each simulation. These refer to the initial mass, profile and nominal Keplerian orbit of the companion galaxy, as well as the disc's Q-profile in the main galaxy.
\section{Results}

\subsection{Evolution of isolated galaxies}

Before inserting the satellite to the system, our first suite of runs shows results of five experiments in which we consider the evolution of the isolated systems for five different Q-values (i.e. changing the asymptotic value $Q_\infty$ in Eq.(\ref{eq:qprofile})), namely $Q=0.5$, $Q=1$, $Q=1.5$, $Q=2$ and $Q=2.5$, (Exp. A1-A5). These experiments also served to check the performance of MAIN, while, fewer body versions of them where repeated using the Barnes and Hut TREE, \citep{Hernquist:2}, yielding quite similar evolution as with MAIN. 

\begin{figure}
\begin{center}
\includegraphics[scale=0.45,angle=0]{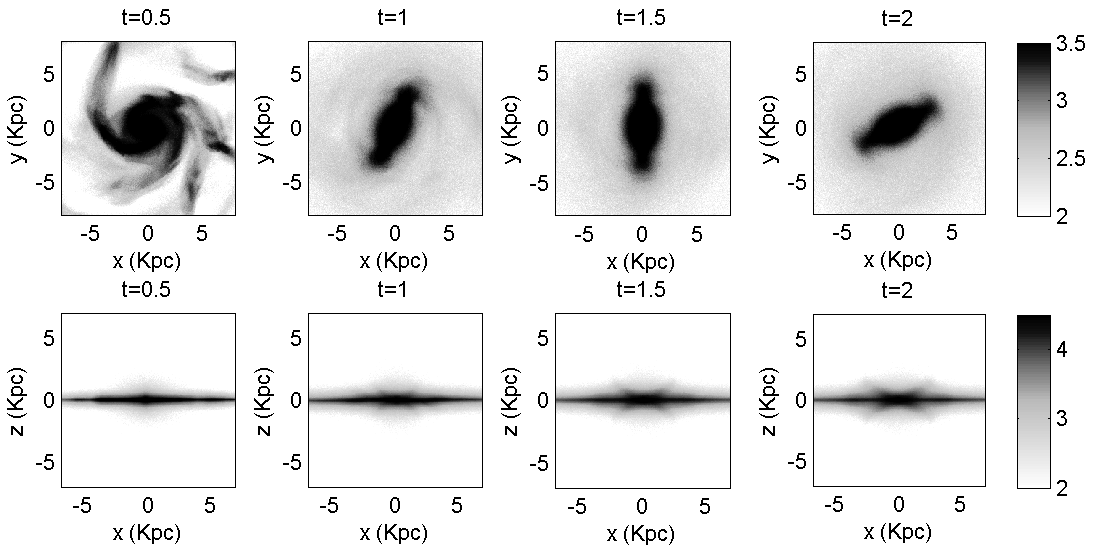}
\end{center}
\caption{Time evolution of the $Q=0.5$, A1 isolated galaxy experiment. Upper row: the projection of the luminous part of the system (bulge+disc) on the $xy$-plane. The color scale represents logarithmic luminosity scale. The times (in Gyr) of each snapshot are indicated on top of each panel. Lower row: Edge-on profiles taken `line-on' i.e. with the bar's major axis coinciding with the horizontal axis of the plot. We observe the characteristic growth of a peanut-shape in a timescale of $\sim 1$Gyr.}
\label{fig:q05iso}
\end{figure}
\begin{figure}
\begin{center}
\includegraphics[scale=0.45,angle=0]{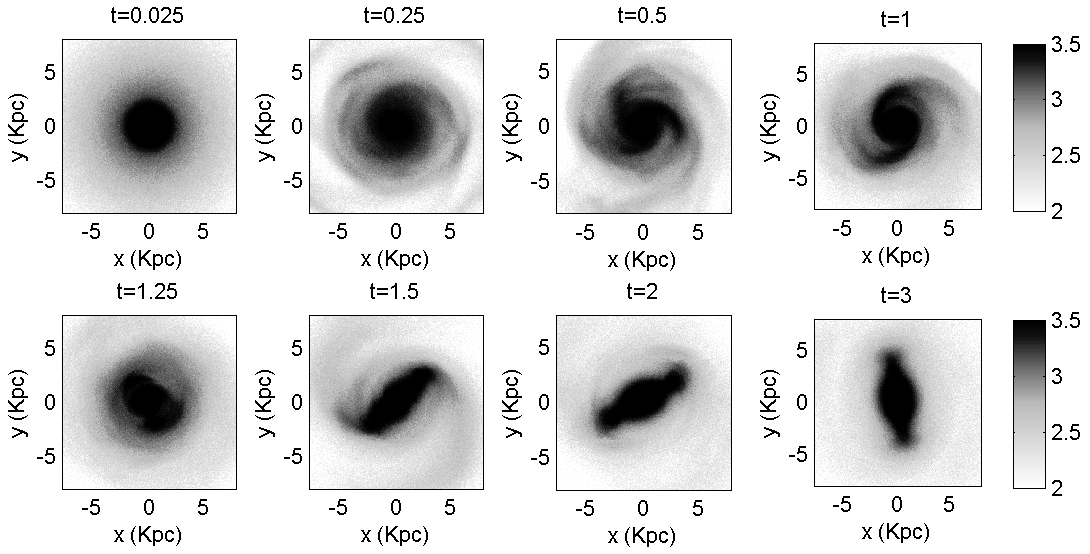}
\end{center}
\caption{Projection on the disc ($xy$) plane of the time evolution of the $Q=1$, A2 isolated experiment. Notation and color scales are as in Fig.\ref{fig:q05iso}. }
\label{fig:q10iso}
\end{figure}
\begin{figure}
\begin{center}
\includegraphics[scale=0.45,angle=0]{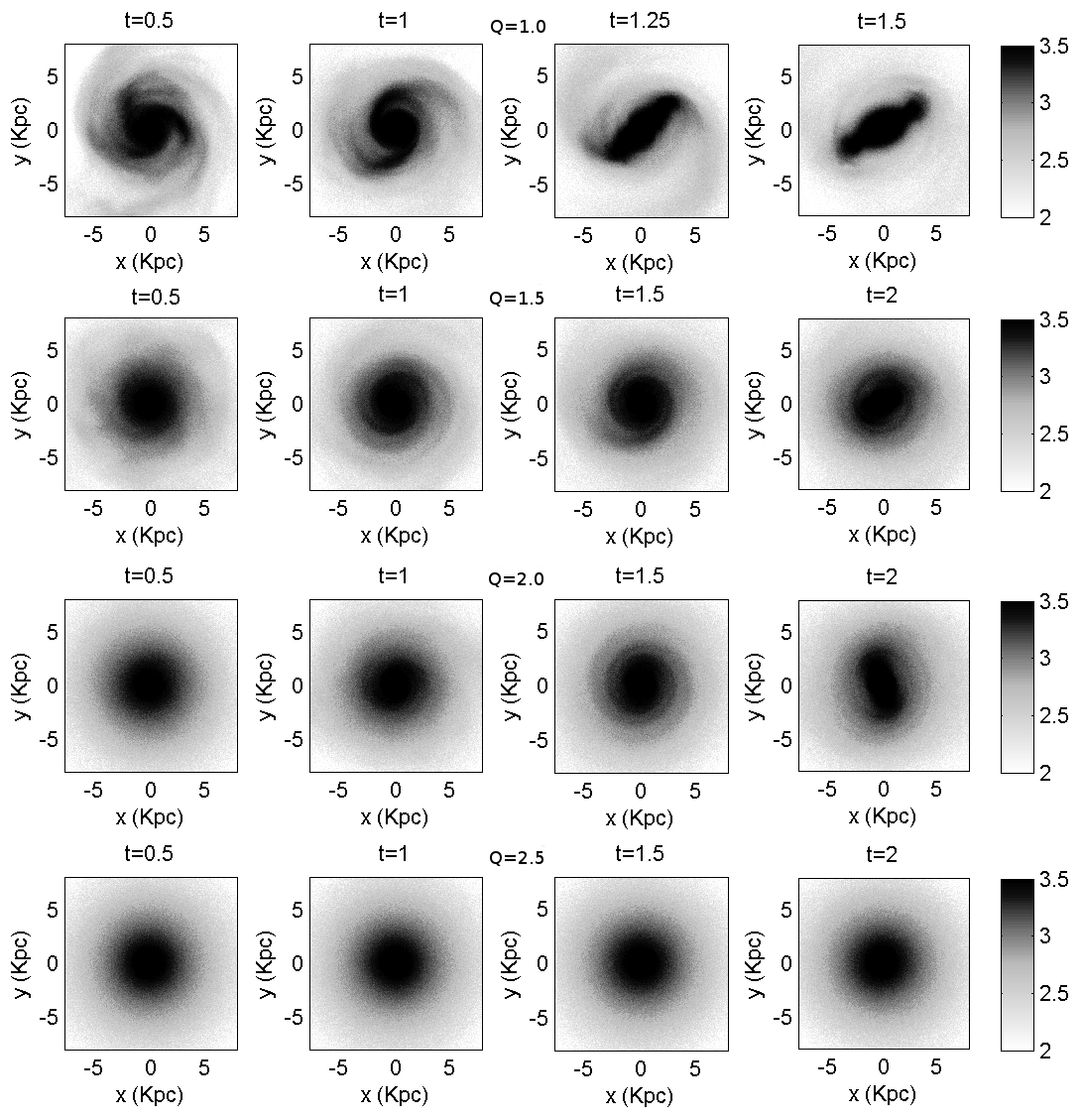}
\end{center}
\caption{Comparative evolution of the isolated galaxy experiments (A2-A5) $Q=1$, $Q=1.5$, $Q=2$ and $Q=2.5$ at four different time snapshots.}
\label{fig:qalliso}
\end{figure}
As expected, experiments with $Q\leq 1$ are prone to both axisymmetric and non-axisymmetric instabilities. In the case of the A1 ($Q=0.5$) experiment (Fig.\ref{fig:q05iso}) the system rapidly develops a characteristic transient $m=3$ instability ($m$ referring here to the Fourier wavenumber of the pattern on the disc with the highest amplitude, see below). However, in less than 500 million years the simulation develops a fast-rotating bar. The latter quickly heats the disc outside it, erasing in less than 1Gyr any important non-axisymmetric feature. After $t=1.5$Gyr there survives nothing more than the bar, which, however, presents itself a secular evolution, leading, in particular, to its thickening and continuous enhancement of peanut shape, as shown in the second row of Fig.\ref{fig:q05iso}.   

The A2 ($Q=1$) isolated experiment presents classical features that have been discussed extensively in the literature since the 70s (see e.g. \citet{Hohl:1, Sellwood:2, Sellwood:3}). Figure \ref{fig:q10iso} gives a summary of the evolution via some characteristic snapshots on the disc plane. After an initial growth of the characteristic 3-arm instability up to $t=0.5$Gyr, bi-symmetric ($m=2$) patterns become dominant in the disc plane, both spiral and bar self-induced modes. The spiral pattern observed at $t=1$Gyr survives unaltered for about two revolutions. However, there is a growing bar mode whose extent on the disc reaches a radius $r\approx 3$Kpc at about $t=1.25$Gyr. Beyond that point the bar mode dominates all accross the disc, producing also recurrent spiral arms for about $1.5$ more Gyrs. These spirals have features of chaotic spiral arms driven by the invariant manifolds of the Lagrangian unstable points, or other unstable periodic orbits, in the end of the bar (see e.g. \citet{Voglis:1, Romero:1, Patsis:1, Tsoutsis:1, Tsoutsis:2, Athanassoula:1}). However, their amplitude decays in time as the bar-spiral activity heats the disc, rendering it less responsive. As a consequence, essentially, all spiral activity beyond the bar ceases after $t=3$Gyr.   

Figure \ref{fig:qalliso} shows now the comparison, at different time snapshots, of four experiments (A2-A5) of isolated galaxies with different initial temperatures, i.e., $Q=1$, $Q=1.5$, $Q=2$ and $Q=2.5$. The main observation is that, even an asymptotic value of $Q_\infty$ as high as 2 does not suffice to completely supress non-axixymmetric self-excited instabilities, although, by increasing $Q$, the instabilities are milder and their onset happens later in time. Thus, comparing the experiments A2 ($Q=1$) with A3 ($Q=1.5$), we observe a clear morphological difference up to $t=2$Gyr, namely, the bar instability is delayed in the A3 ($Q=1.5$) experiment, allowing, instead, for spiral modes inside the disc's inner 5Kpc to survive for longer. In the A3 ($Q=1.5$) experiment, we find that these modes survive nearly for five revolutions, showing some recurrent decrease and increase in the amplitude, but being always clearly identifiable in a Fourier map of the surface density (see below), with $m=2$ amplitudes $0.1\sim 0.2$. These modes, however, evolve into a growing bar mode after $t\simeq 1.8$Gyr. Similar spiral modes were observed, but with smaller amplitudes, in the A4 ($Q=2$) experiment, which, however, still exhibits a growing bar mode whose onset is a little later than in the case of the A3 ($Q=1.5$) experiment, i.e., at $t=2$Gyr. Finally, all such non-axisymmetric modes seem to be suppressed in the A5 ($Q=2.5$) experiment, whose disc remains essentially featureless up to the end of the integration, i.e., at $t=3$Gyr. 

\subsection{Evolution with companion flybys: mass window}
We now pass to our main corp of simulations including companion galaxies. The A2 ($Q=1$) isolated galaxy experiment excites self-modes quite rapidly, which are observed to dominate the evolution of the system with or without companion. Instead, in the A3 ($Q=1.5$) experiment we find that the self-induced and the companion-induced modes grow in antagonism. In our first set of experiments, we fix the main galaxy to be the $Q=1.5$ system, and we set the companion at a relative orbit with initial quasi-Keplerian elements $a=27$Kpc, $e=0.4$ $I=30^\circ$. This implies a first pericentric passage at $\approx 15$Kpc. We then made several experiments (Exp. B3-B5), by progressively increasing the satellite's initial mass from $M_s=5\times 10^9M_\odot$ to $M_s=2\times 10^{10}M_\odot$. The snapshots shown in Figs.\ref{fig:q15_1e9}, \ref{fig:q15_1e10}, and \ref{fig:q20_2e10} refer to the satellite mass choices $M_s=5\times 10^9M_\odot$, $10^{10}M_\odot$ and $2\times 10^{10}M_\odot$ respectively. These correspond to initial mass ratios 1:10, 1:5 and 2:5, respectively, of the satellite's over the disc's mass. 

\begin{figure}
\begin{center}
\includegraphics[scale=0.317,angle=0]{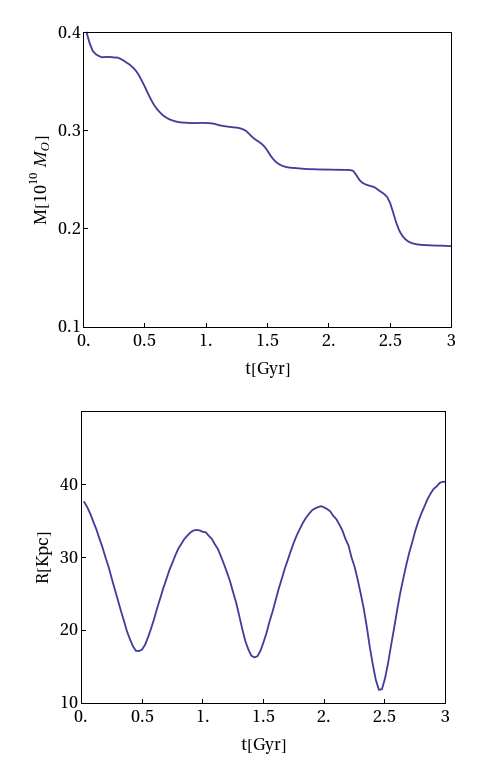}
\includegraphics[scale=0.3,angle=0]{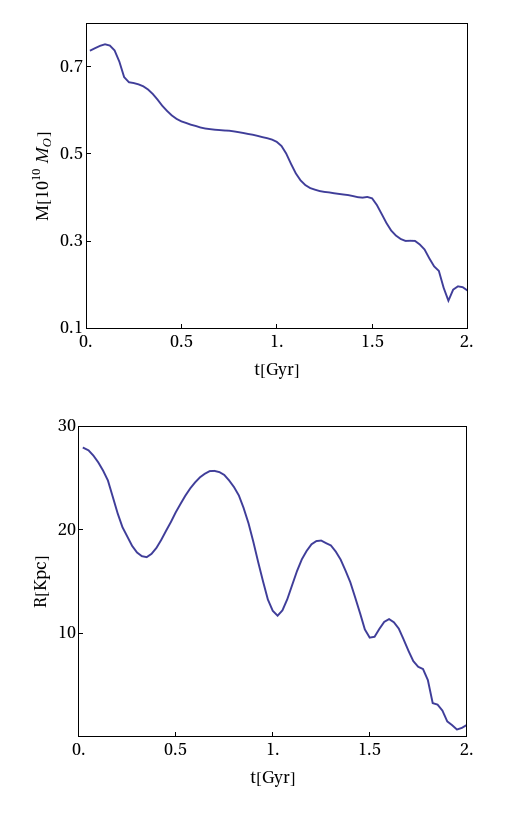}
\includegraphics[scale=0.305,angle=0]{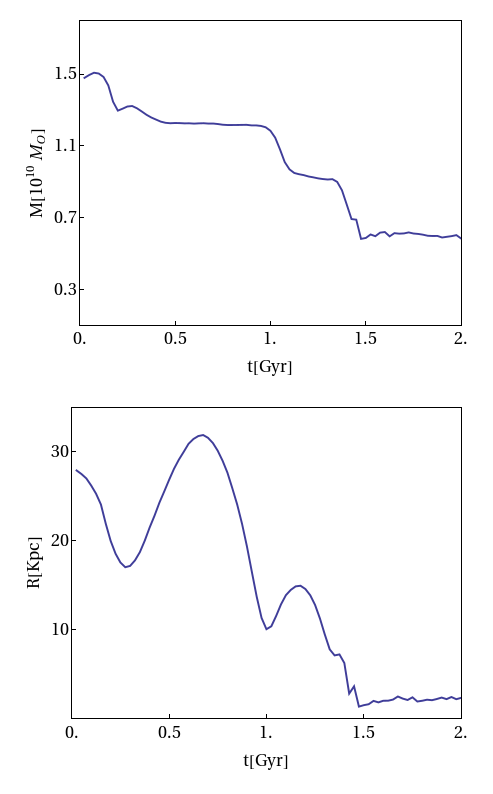}
\end{center}
\caption{Top three panels: time evolution of the mass bounded to the satellite (the quantity $M_{s,r<4Kpc}$, see text) in the three experiments with initial disc $Q=1.5$ and initial satellite to disc mass ratios 1:10 (left), 1:5 (center), 2:5 (right), (Exp. B3-B5). Low three panels: time evolution of the satellite - disc (center of mass) distance in the same experiments.}
\label{fig:massloss}
\end{figure}
Figure \ref{fig:massloss} shows an overview of the satellite's orbital decay of the above three experiments. In order to compute approximately the orbit of the center of mass of the satellite, we identify in the initial conditions, $10\%$ of the most satellite's gravitational bound bodies. These bodies remain bound all the way up to the merger of the satellite with the disc. Then we identify the numerical trajectory of the satellite with the trajectory of the center of mass of these most bound bodies. This allows to estimate a mass loss rate for the satellite during the orbital decay. To this end, we compute the quantity $M_{s,r<4Kpc}(t)$, i.e. the total mass of the satellite's bodies which remain within a distance less than 4Kpc from the satellite's center of mass at the time t. The limit of 4Kpc is rather arbitrary, but quite convenient in numerical computations in comparison with the cumbersome exact computation of the satellite's Hill sphere.
With the above assumptions, the top three panels of Fig. \ref{fig:massloss} show the time evolution of the mass $M_{s,r<4Kpc}(t)$ for the three experiments, while the lower three panels show the distance between the satellite's center of mass and the main galaxy disc's center of mass. We notice immediately that the mass loss rate $\left | \dot{M}_{s,r<4Kpc}(t) \right |$ is not uniform in time, but increases at every approach of the satellite near pericentric passage. This is due both to the stronger tidal field and to the larger dark matter halo density close to the center of the disk. Most importantly, the mass  decreases significantly in all the experiments, nearly at a factor of 2 with respect to $M_{s,r<4Kpc}(0)$ after the second pericentric passage. As a result, in, say, the middle column of Fig. \ref{fig:massloss}, we find $M_{s,r<4Kpc}=4\times10^9M_\odot$ after the second pericentric passage, and $M_{s,r<4Kpc}=3\times10^9M_\odot$ after the third pericentric passage, which corresponds to satellite to disk mass ratios of 1:14 and 1:18 respectively, i.e. significantly lower than the initial ratio (1:5). However, as shown below, these mass-depleted satellites, are able to excite spiral structure in the main galaxy's disc. We thus conclude that effective mass ratios of only a few percent are enough to excite spiral structure, as noted already in \citet{Byrd:1}, but the satellite progenitors must be a 2-3 times heavier initially in order to reach this mass level at the closest encounters with the disc via the repeated flyby mechanism.

Keeping the above remarks in mind, we now discuss separately the three experiments, referring to them by the initial mass ratio, i.e. the '1:10', '1:5' and '2:5' experiment. 

In the case of the B3 (1:10) experiment (Fig.\ref{fig:q15_1e9}) we found no important dynamical influence of the satellite on the disc even at the closest pericentric passages. As shown in the first two panels of Fig.\ref{fig:q15_1e9}, the satellite's orbit in-spirals towards the center of the main galaxy at a rate determined by the dynamical friction with the dark halo. However, this rate is low enough so that the satellite galaxy does not merge with the disc during the whole integration period up to $t=3$Gyr. However, at $t=2$Gyr the $Q=1.5$ disc develops a self-excited bar instability, which determines the disc's evolution at later stages. Comparing the $t=2$Gyr snapshots of the ($Q=1.5$) experiment with or without satellite (last panel in Fig.\ref{fig:q15_1e9} with last panel of second row of Fig.\ref{fig:qalliso}) we see that the two bars are quite similar, the former one being only slightly more elongated than the latter. Since the orbit of the satellite nearly preserves its inclination at $30^\circ$, i.e., not very far from the equator, the additional elongation can be attributed to the time-averaged nearly-equatorial tide exerted by the satellite upon the main galaxy.   

A quite distinct evolution takes place, instead, when the initial satellite/disc mass ratio is set to B4 (1:5, Fig.\ref{fig:q15_1e10}). In this case, the satellite makes three successive pericentric passages at the times $\approx$ $t=0.7$Gyr, $t=1.3$Gyr and $t=1.8$Gyr. Everyone of them is associated with an excitation, of progressively higher amplitude, of a $m=2$ mode in the main galaxy's disc. This has an oval bar-like form in the inner parts of the disc (up to about $4$Kpc), and a spiral form beyond that distance, which can be traced in the Fourier maps of the surface density accross the whole disc (see subsection 4.3 below).  The second and third panels of the last row of Fig.\ref{fig:q15_1e10} show these $m=2$ features at two time snapshots shortly after the second and third satellite's pericentric flyby. These figures show the striking difference between companion-induced and self-induced modes (compare with the second row of Fig.\ref{fig:qalliso}). As detailed in subsection 4.4, the companion-induced spiral arms have $m=2$ normalized Fourier amplitudes growing with the distance from the center, which are visible across nearly the whole optical disc (at distances $\sim 10$Kpc), forming also tidal tails all the way to the instantaneous pericenter of the satellite's orbit. Measuring the spiral pattern speed we find values between $20$ and $30$Km$s^{-1}Kpc^{-1}$, implying periods $200\sim 300$ Myr. On the other hand, depending on its distance from the main galaxy, the mean rotation period of the satellite evolves from less than $10$Km$s^{-1}Kpc^{-1}$ initially to about $12$Km$s^{-1}Kpc^{-1}$ at the third pericentric passage. Thus, the tidally induced spiral modes on the disc are not in 1:1 spin resonance with the orbit of the satellite. Instead, as evidenced in subsection 4.4, these spiral arms have the features of density waves, whose amplitude's decay over time is essentially compensated by a repeated amplification at the successive companion flybys. In Fig.\ref{fig:q15_1e10} the spiral waves become dominant in the disc shortly after the second companion's pericentric passage, at $t=1.3$Gyr, and survive in a grand-design form nearly continuously in time up to the fourth companion's pericentric passage, at $t=2.3$Gyr. The spiral waves make about four revolutions within this time interval. Beyond this time, however, the fourth satellite's pericentric passage is at a distance close enough for the satellite to be tidally disrupted (last column in Fig.\ref{fig:q15_1e10}). The fall of the satellite into the disc is itself not without consequences. Besides thickening the disc, we find that it generates a strong bar instability, which leads to a formation of a rapidly rotating bar. Note that this is a late companion-induced effect. Actually, at earlier times, the presence of the satellite {\it suppressed} the growing of the bar self-mode in the same galaxy; in the corresponding isolated galaxy the onset of the bar instability was already observed at $t=1.8$Gyr. 
\begin{figure}
\begin{center}
\includegraphics[scale=0.4,angle=0]{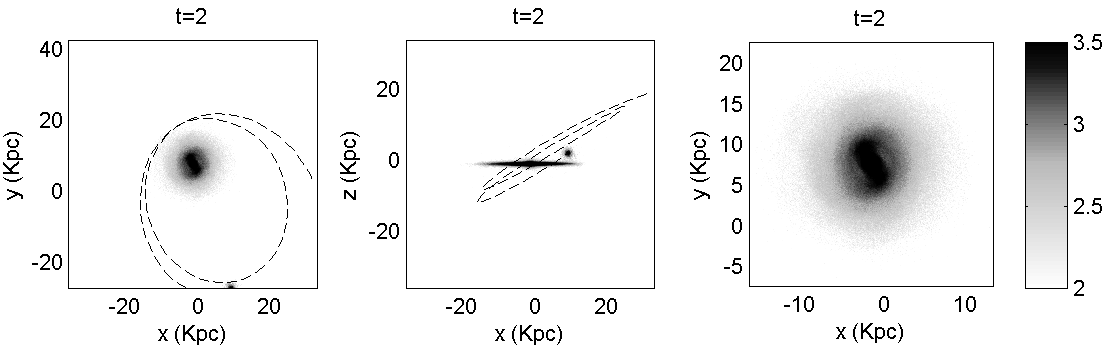}
\end{center}
\caption{Simulation with the main galaxy of disc's initial Q-value $Q=1.5$ and satellite with initial mass $M_s=5\times 10^{9}$ in initial quasi-Keplerian orbit of elements $a=27$Kpc, $e=0.4$, $I=30^\circ$, (Exp. B3). The dashed red curve indicates the orbit of the satellite up to the time $t=2$.  This is superposed to the image of the main galaxy at the same time, projected on the xy-plane (left) or xz-plane (right). A greater zoom to the xy-plane is shown in the right panel.}
\label{fig:q15_1e9}
\end{figure}
\begin{figure}
\begin{center}
\includegraphics[scale=0.4,angle=0]{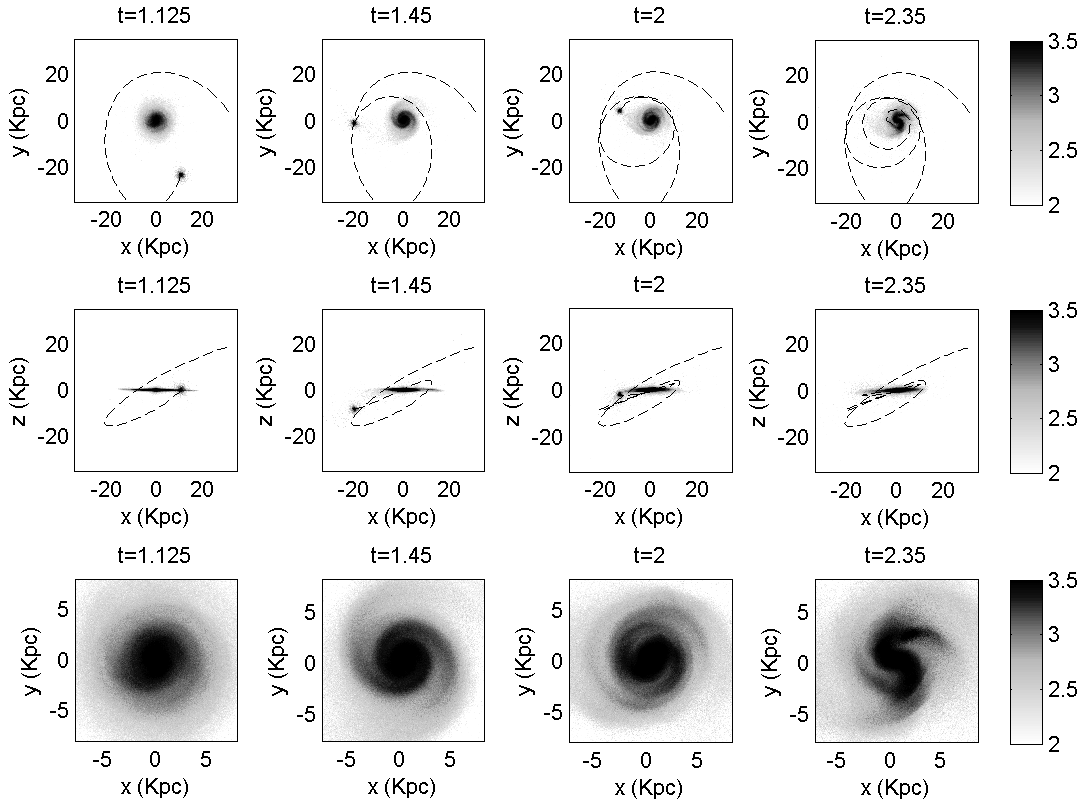}
\end{center}
\caption{Snapshots of the time evolution of the experiment with the main galaxy of disc's initial Q-value $Q=1.5$ and satellite with initial mass $M_s=10^{10}$ in initial quasi-Keplerian orbit of elements $a=27$Kpc, $e=0.4$, $I=30^\circ$, (Exp. B4). Top and middle rows: projection of the evolution of the system in the xy- and xz-plane. Bottom row: the disc evolution in greater zoom, projection on the xy-plane.}
\label{fig:q15_1e10}
\end{figure}
\begin{figure}
\begin{center}
\includegraphics[scale=0.4,angle=0]{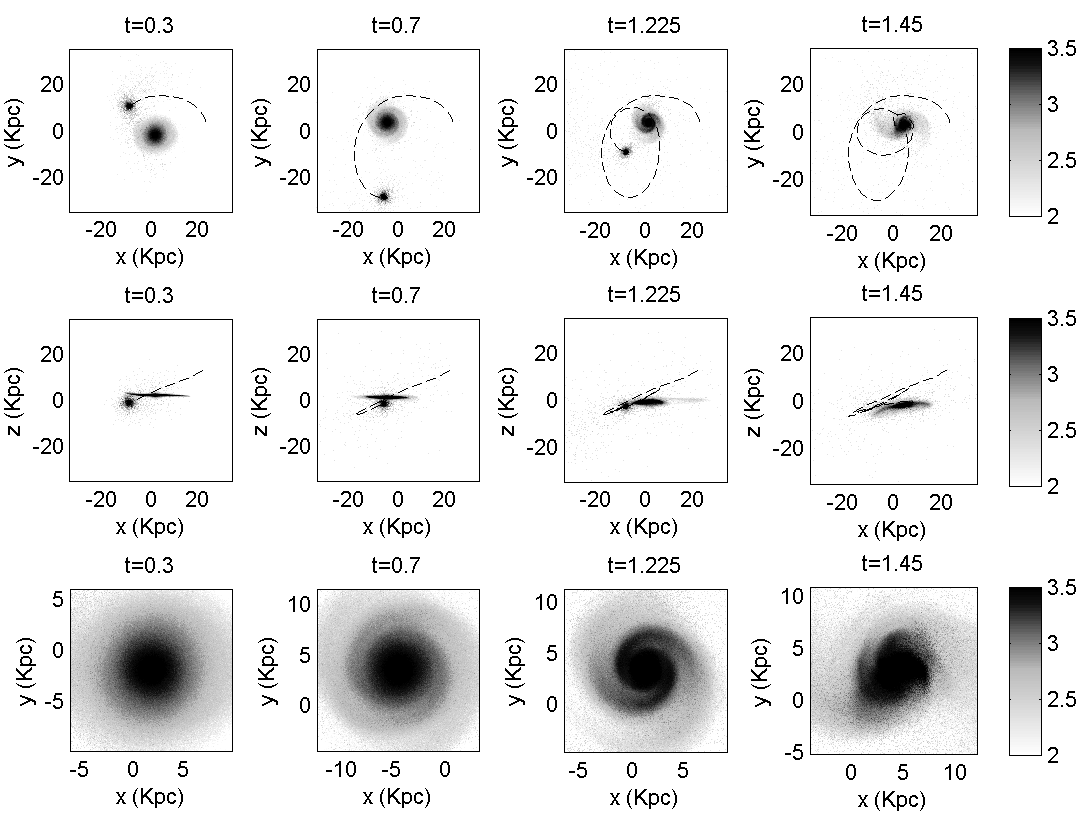}
\end{center}
\caption{Same as in Fig.\ref{fig:q15_1e10}, but increasing the satellite mass to $M_s=2\times 10^{10}$., (Exp. B5)}
\label{fig:q20_2e10}
\end{figure}

Figure \ref{fig:q20_2e10} shows, now, what happens if the satellite's initial mass $M_s$ is $2\times 10^{10}M_\odot$, i.e., an initial satellite/disc mass ratio 2:5, (B5). The orbital decay rate is now larger and the satellite suffers tidal disruption and merges with the main disc at the third pericentric passage, at $t=1.4$Gyr. The second pericentric passage, at $t=1$Gyr, generates again a conspicuous spiral wave, which, however, survives for less than two revolutions before the satellite's final fall into the disc. 

In conclusion, we find that the satellite initial mass in these experiments has to be in a rather narrow range of values in order to generate spiral waves able to survive for several revolutions in the main galaxy's disc, before the satellite's orbital decay makes the system to radically change morphology. Initial satellite to disc mass ratios around 1:5 are needed in our specific initial conditions. Although it seems probable that the result depends also on the eccentricity and inclination of the initially considered quasi-Keplerian satellite orbit, for a fixed initial satellite distance of $\sim 25Kpc$, a mass ratio 1:3 appears to lead, in general, to an orbital decay so fast, that it essentially destroys the effectiveness of the repeated flyby mechanism. It should be stressed, however, that the range of our hereby reported mass windows refers to companion's orbits regulated so that the pericentric passages associated with the excitation of spiral structure occur at distances representing the outskirt of the main galaxy's disc ($10$ to $15$Kpc in our experiments). Instead, quite different results are found if the companion penetrates the disc at shorter distances. For example, the first experiment reported in \citep{Donghia:1} uses parameters similar to ours, but with a companion's pericentric passage at about 2Kpc from the center of the disc. This results in a quite different evolution, namely, some initially excited two-arm spirals evolve into a ring structure propagating outwards across the main galaxy's disc after the companion's first flyby through the disc. On the other hand, in both types of experiments the disc has a Q-profile with values well above unity, and quite massive companions, i.e., of masses of the order of $10^{10}M_\odot$ are needed to generate an immediate response of the disc.   
\clearpage
\subsection{Dependence on the disc's temperature}
\begin{figure}
\begin{center}
\includegraphics[scale=0.45,angle=0]{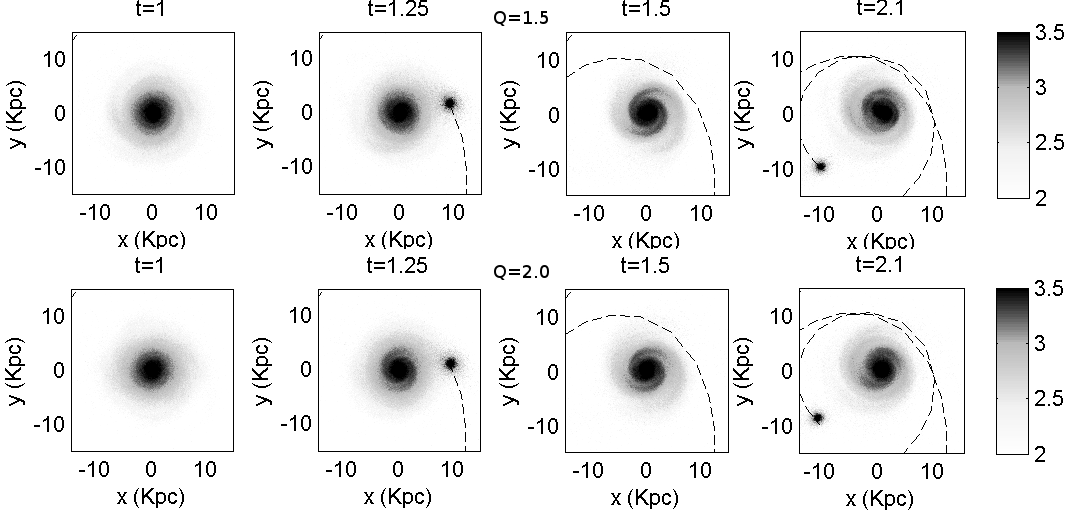}
\end{center}
\caption{Excitation of $m=2$ spiral waves by companion flybys in experiments B4 and B6 in which the satellite has mass $M_s=10^{10}M_\odot$ (satellite to disc mass ratio 1:5), when the satellite is initially placed in a quasi-Keplerian orbit of elements $a=27$Kpc, $e=0.4$, $I=30^\circ$, and the main galaxy's disc has an initial asymptotic Q-value $Q=1.5$ (upper row), or $Q=2$ (lower row).}
\label{fig:q1520_1e10}
\end{figure}

Besides the orbital decay rate of the companion (depending on its mass $M_s$ and on the dark matter density profile $\rho_h$), another factor which plays a role in the development of the companion-induced structures is the initial temperature (or Q-profile) of the main galaxy's disc. A higher temperature tends to suppress the onset self-excited modes on the disc, while, up to a certain temperature limit, the tides exerted by the companion can still be effective in exciting large-amplitude spiral waves. Figure  \ref{fig:q1520_1e10} provides the relevant information. The upper row corresponds to the same experiment as in Fig.\ref{fig:q15_1e10}, in which we start with a $Q=1.5$ disc, (Exp. B4), while the lower row shows an experiment in which we start with a $Q=2$ disc, (Exp. B6). The satellite's orbit is quite similar in both experiments, and successive pericentric passages take place in both at nearly equal times. Two main remarks are: 

i) The satellite is able to exite spiral waves in the $Q=2$ disc, even while in the isolated galaxy simulations this disc was found quite non-responsive to the growth of spiral self-modes (third row in Fig.\ref{fig:qalliso}). This is in agreement with the results of experiments based on parabolic (rather than bound) companion orbits,  \citep{Struck:1, Oh:2}.

ii) At $t=2.1$Gyr, the $Q=1.5$ disc exhibits the onset of a self-excited bar-instability, while the $Q=2$ disc still exhibits a companion-excited spiral wave. Thus, in the B4 ($Q=1.5$) experiment, the growth of the self-excited modes prevails the growth of the companion-induced modes at a time before the companion's final merger. On the contrary, in the B6 ($Q=2$) experiment, the companion-induced modes prevail in the disc over the whole time interval from the first pericentric passage up to the companion's final merger. 

\begin{figure}
\begin{center}
\includegraphics[scale=0.42,angle=0]{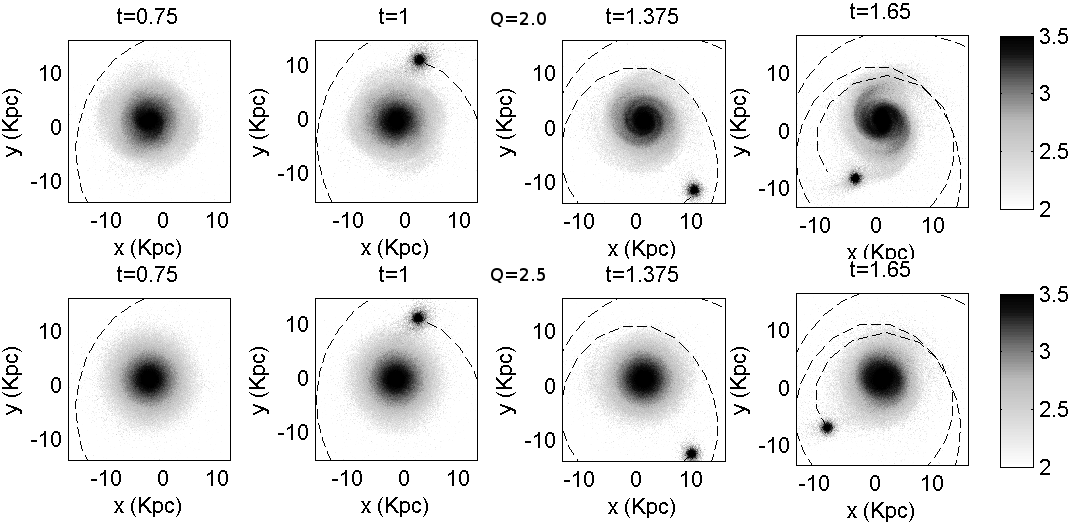}
\end{center}
\caption{Upper row: excitation of $m=2$ spiral waves in the experiment with $M_s=10^{10}M_\odot$ (mass ratio 1:5), and quasi-Keplerian orbit elements $a=20$Kpc, $e=0.4$, $I=30^\circ$, and a main galaxy's disc initially with $Q=2.0$. Lower row: same experiment, but with initial disc Q-value $Q=2.5$. Now, no spiral waves are excited by the companion flybys on the main galaxy's disc., Exp. (B7-B8)}
\label{fig:q2025_1e10}
\end{figure}

On the other hand, we have performed several tests, using also different satellite orbits, but failed to excite significant spiral structure in any experiment in which the main galaxy's disc was equal to $Q=2.5$, (Exp. B8). Fig.\ref{fig:q2025_1e10} shows two experiments, in which the satellite's orbit is regulated to a somewhat lower major semi-axis than in previous plots, i.e. with the initial elements $a=20$Kpc, $e=0.4$, $I=30^\circ$). In both experiments (B7-B8) of Fig.\ref{fig:q2025_1e10}, the first pericentric passage is now at only four exponential disc lengths ($r\approx 12$Kpc). The spiral structure generated in the B7 ($Q=2$) experiment (upper row) has very similar features with the one found in the $Q=2$ experiment of Fig.\ref{fig:q1520_1e10}. Yet, the same satellite of such mass and orbit leaves nearly featureless the disc of the B8 ($Q=2.5$) experiment, even after three pericentric passages.

Thus, we conclude that values of $Q$ around $Q=2$, i.e., relatively hot stellar discs, are both susceptible to and favoring the longevity of companion-induced spiral modes. At lower values of Q, the disc tends to becomes dominated by self-induced modes outstanding the companion's effect before the latter's infall, while, at higher values of Q, the disc becomes non-responsive to the companion's impulses even at the companion's closest pericentric passages.


\subsection{Features of the excited spirals}
We now focus on some features of the spirals excited by the satellite as observed in the previous figures. We studied several examples of such spirals, and they all share a number of common features, best revealed through a spectral analysis of the surface density and potential in the main galaxy's disc at the corresponding snapshots when these spirals appear more conspicuous. 

We focus below on one such example, corresponding to the spiral pattern shown in the upper row of Fig.\ref{fig:q2025_1e10}, Exp. B8. The projected face-on surface density of the disc $\Sigma_d(R,\phi)$, expressed in polar co-ordinates centered around the disc-bulge center of mass, is Fourier-decomposed as:
\begin{equation}\label{eq:foursig}
\Sigma_d(R,\phi)=\Sigma_{d,0}(R) 
+ \sum_{m=1}^\infty \bigg(A_m(R)\cos(m\phi)+B_m(R)\sin(m\phi)\bigg)
\end{equation}
where
$$
\Sigma_{d,0}(R)={1\over 2\pi}\int_0^{2\pi}\Sigma_d(R,\phi)d\phi,~~~
A_m(R)={1\over \pi}\int_0^{2\pi}\Sigma_d(R,\phi)\cos(m\phi)d\phi,~~~
B_m(R)={1\over \pi}\int_0^{2\pi}\Sigma_d(R,\phi)\sin(m\phi)d\phi,~~m=1,2,\ldots
$$
We call amplitude $C_m(R)$ and phase $\phi_m(R)$ of the mth mode the quantities
\begin{equation}\label{eq:ampphase}
C_m(R) = {\left(A_m^2(R)+B_m^2(R)\right)^{1/2} \over \Sigma_{d,0}(R)},~~~
\phi_m(R) = \tan^{-1}(B_m(R)/A_m(R))
\end{equation}
\begin{figure}
\begin{center}
\includegraphics[scale=0.3,angle=0]{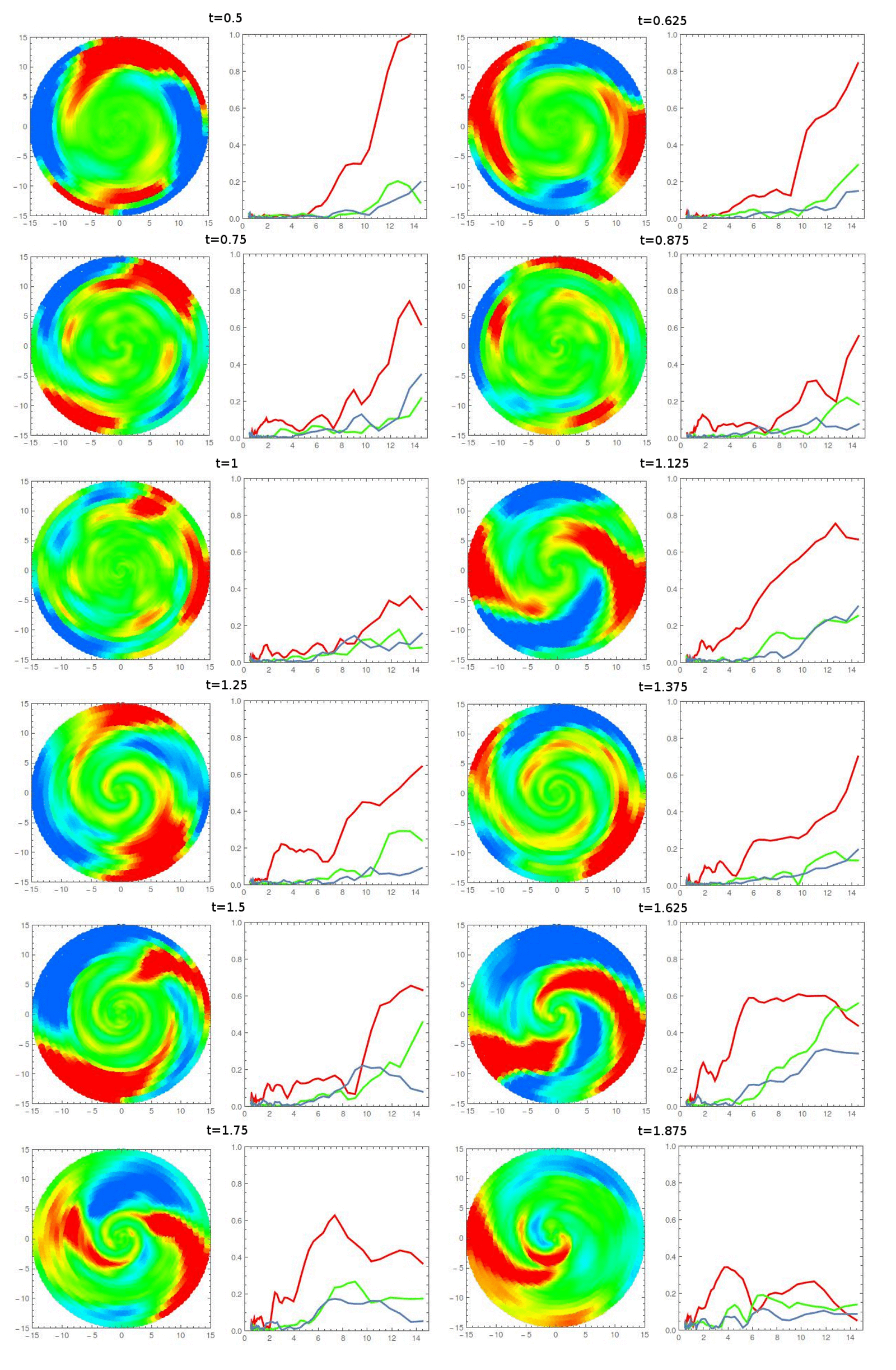}
\end{center}
\caption{Twelve double panels representing different snapshots (at the indicated times) of the experiment B8 with initial data $Q=2$, $M_s=10^{10}M_\odot$, $a=20$Kpc, $e=0.4$, $I=30^\circ$. In each snapshot, the left panel shows a color map of the quantity $D_s(R,\phi)$ (smoothed excess density, Eq.(\ref{eq:exden})). The color scale ranges from blue, for the value $D(R,\phi)=-0.5$ or smaller, to red, for $D(R,\phi)=0.5$ or larger. The right panel shows the Fourier amplitudes $C_2(R)$ (red), $C_3(R)$(green), and $C_4(R)$(blue), see Eq.(\ref{eq:ampphase}), as a function of the distance $R$ in Kpc from the disc's center.}
\label{fig:fourier}
\end{figure}

Figure \ref{fig:fourier} shows a Fourier analysis of the disc's projected surface density in our example. Based on the Fourier transforms of Eqs.(\ref{eq:foursig}) we compute first a smoothed projected surface density function, obtained by the order-10 truncated Fourier series
\begin{equation}\label{eq:sigsm}
\Sigma_s(R,\phi)=\Sigma_{d,0}(R) 
+ \sum_{m=1}^{10} \bigg(A_m(R)\cos(m\phi)+B_m(R)\sin(m\phi)\bigg)
\end{equation}
In each double panel, now, of Fig.\ref{fig:fourier}, referring to a different time snapshot, the left panel shows in color scale the `smoothed excess density', i.e. the quantity
\begin{equation}\label{eq:exden}
D_m(R,\phi) = {\Sigma_s(R,\phi)-\Sigma_{d,0}(R) \over \Sigma_{d,0}}~~~. 
\end{equation}
The computations are done after arranging the particles in a polar grid, with 50 logarithmically spaced radial bins from $R_1=0.1$Kpc to $R=15$Kpc, and 180 azimuthal bins from $\phi=0$ to $2\pi$. Also, the right panel shows in red, green and blue, the amplitudes $C_2(R)$, $C_3(R)$ and $C_4(R)$ respectively. The figure shows data from twelve different time snapshots in the interval from $t=0.5$Gyr (shortly after the first satellite's pericentric passage, which occurs at $t=0.37$Gyr), to $t=1.9$Gyr, the merger time of the satellite with the main galaxy. 

\begin{figure}
\begin{center}
\includegraphics[scale=0.8,angle=0]{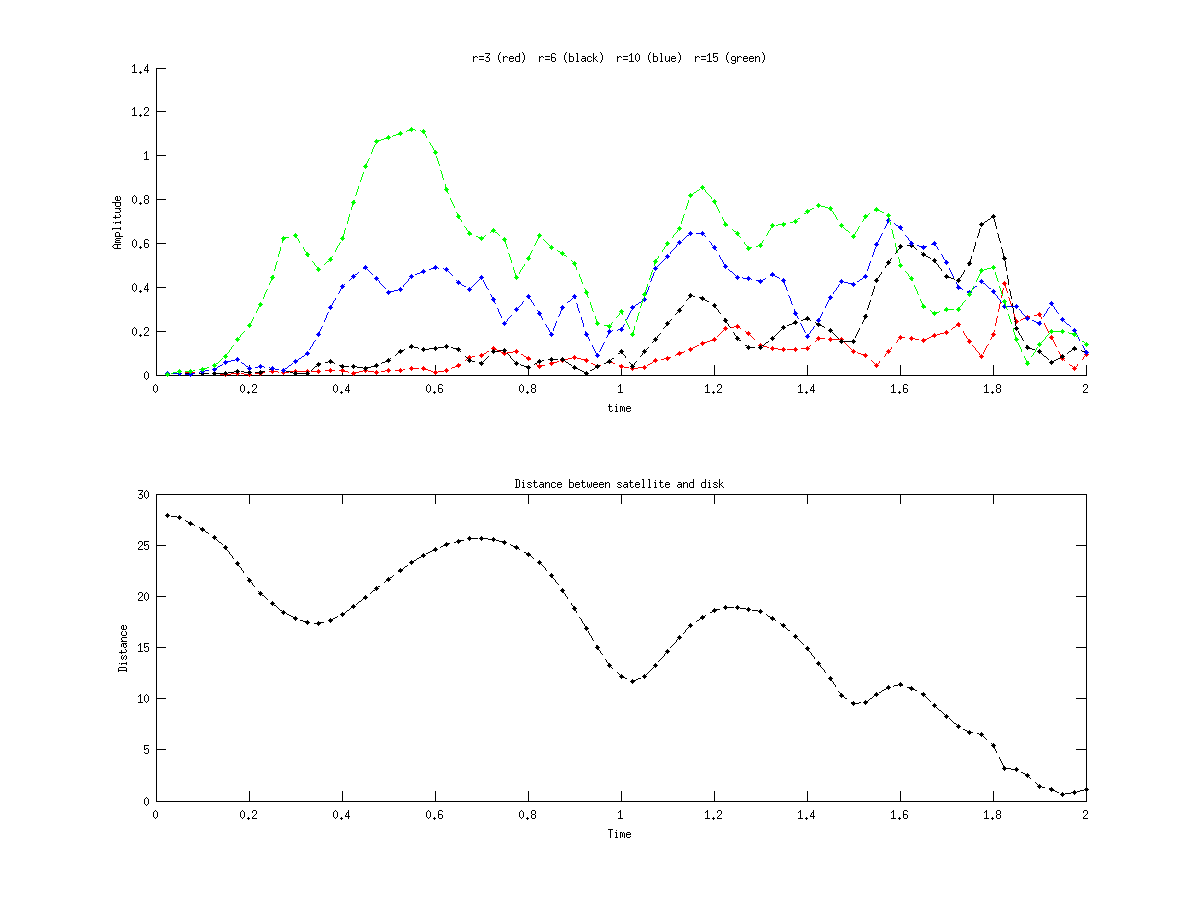}
\end{center}
\caption{Time evolution of the $m=2$ Fourier mode amplitude ($C_2(R)$, Eq.(\ref{eq:ampphase}) at four different radii with respect to the disc's center, namely $R=3,6,9$ and $12$Kpc, compared to the time evolution of the distance between the center of the disc and the center of the satellite.}
\label{fig:ampr}
\end{figure}

As evident in Fig.\ref{fig:fourier}, a conspicuous $m=2$ mode is nearly always present accross the disc as a response to the satellite's tide, whose leading Fourier terms are also $m=2$ (the $m=1$ terms correspond to the acceleration of the main galaxy's center of mass with respect to the satellite's center of mass; hence, such terms are absorbed in a reference frame centered on the main galaxy's center of mass). At the outskirts of the disc, the tidally induced $m=2$ amplitude can be as high as unity. Also, many snapshots (e.g. all snapshots from $t=0.5$ to $t=1$Gyr but also $t=1.25$ or $t=1.5$Gyr) are suggestive of a threshold radial distance from the disc's center $R\approx 2$Kpc, below which the disc remains essentially non-responsive to the tide, while, beyond that distance, the amplitude of the $m=2$ mode grows roughly linearly with $R$. This threshold distance may be related to the exponential scale distance $R_q$ of the initial Q-profile of the disc (Eq.(\ref{eq:qprofile})). However, this picture changes after the appearance of the spiral density wave on the disc, in particular at times after $t=1$, i.e. after the second satellite's pericentric passage. It should be noted that the dynamical response of the disc to the tide depends not only on the strength of the tide but also on the internal matter distribution within the disc. Also, in the inner parts of the disc, in particular, we observe an oscillation in time of the amplitude $C_2(R)$, which is connected to the orbital period of the satellite. This phenomenon is more evident in Fig.\ref{fig:ampr}, showing the time variation of the $m=2$ amplitude $C_2(R)$ at four different radii, namely $R=3,6,9$ and $12$Kpc, as correlated with the satellite's orbit, i.e., distance between the satellite's and disc's centers of mass. Besides the obvious time oscillations of the $m=2$ amplitude found at all distances, we observe, most notably, the presence of a phase difference between the satellite's pericentric passages and subsequent growth of the disc's $m=2$ mode; actually, the pericentric passages themselves are always very close to {\it minima} of the $m=2$ amplitude. Nevertheless, we note the general tendency of growth of the $m=2$ mode on the disc as the satellite's orbit decays closer and closer to it, until, eventually, these modes decay after the merger of the satellite with the main galaxy. 

\begin{figure}
\begin{center}
\includegraphics[scale=0.42,angle=-90]{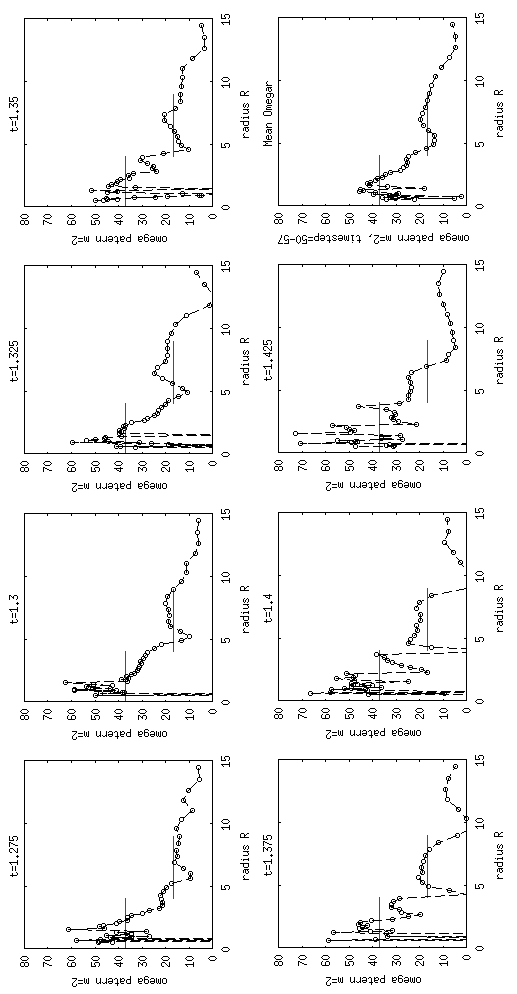}
\end{center}
\caption{The quantity $\omega_2(R,t)$ (Eq.(\ref{eq:omedphi})) as a function of $R$ for seven different time snapshots in the interval $1.275$Gyr$\leq t \leq 1.425$Gyr, shortly after the satellite's second pericentric passage. The last panel represents a `stack plot' (addition and division by 7) of the seven previous plots. The red lines show the mean value $\Omega_p=<\omega_2>$ as derived from the stack plot in the intervals $1.5$Kpc$\leq R\leq 3.5$Kpc (inner red line) and $4$Kpc$\leq R\leq 10$Kpc (outer red line).}
\label{fig:meanome1}
\end{figure}

\begin{figure}
\begin{center}
\includegraphics[scale=0.42,angle=-90]{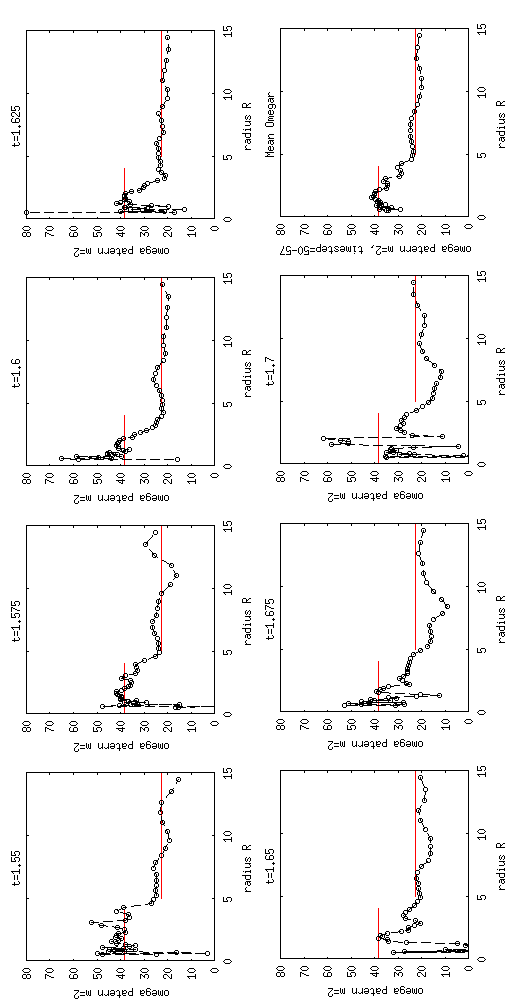}
\end{center}
\caption{Same as in Fig.\ref{fig:meanome1}, but for the time interval $1.55$Gyr$\leq t \leq 1.7$Gyr, shortly after the third satellite's pericentric passage.}
\label{fig:meanome2}
\end{figure}

Tracking, now, besides the amplitudes, the phases $\phi_m(R)$ of the various Fourier modes, Eq.(\ref{eq:ampphase}) allows us to define the pattern speeds, or frequencies, connected with the propagation of the various modes on the disc. We focus again on the $m=2$ mode. Figures \ref{fig:meanome1} and \ref{fig:meanome2} provide the relevant information in the case of the experiment in Fig.\ref{fig:fourier}. These figures refer to two different time intervals, namely $t=1.275$ to $t=1.425$ (Fig.\ref{fig:meanome1}) and $t=1.55$ to $t=1.7$ (Fig.\ref{fig:meanome2}). The first interval is shortly after the second satellite's pericentric passage (at $t=1$, see Fig.\ref{fig:ampr}), while the second interval is after the satellite's third pericentric passage (at $t=1.5$). In both these intervals we have conspicuous spiral arms in the disc. The spirals complete in total two revolutions in the time interval from $t=1.2$ to $t=1.7$, while, the time span in each of the Figs.\ref{fig:meanome1} and \ref{fig:meanome2} represents half of the first and second  revolution respectively. Consider now, two nearby in time values of the phase $\phi_2$ at fixed distance $R$, namely $\phi_2(R,t)$ and $\phi_2(R,t+\Delta t)$. The instantaneous rotation frequency of the $m=2$ mode at the distance $R$ can be approximated by:
\begin{equation}\label{eq:omedphi}
\omega_2(R,t) \simeq {\phi_2(R,t+\Delta t)-\phi_2(R,t)\over \Delta t} 
\end{equation}
A pattern speed for the $m=2$ mode is well defined in a certain domain $R_1<R<R_2$ accross the disc provided that $\omega_2(R,t)$ remains nearly constant, for fixed $t$ and all $R$ within the domain, and also for fixed $R$ and $t$ varying within the considered time interval. Implementing these criteria in the case of Figs.\ref{fig:meanome1} and \ref{fig:meanome2}, we can distinguish several approximate `plateaus' of the curve $\omega(R,t)$ yielding local values of the $m=2$ pattern speed in both time intervals. The most conspicuous plateau is observed in the second time interval (Fig.\ref{fig:meanome2}), which is also connected to a higher amplitude of the $m=2$ mode excited after the third pericentric passage of the satellite. This plateau, evident at all times, extends from $R\approx 4$Kpc and beyond throughout the whole optical disc. The last panel in Fig.\ref{fig:meanome2} represents a mean 'stack plot' in time, i.e. sum and division by seven of the seven previous plots, which represent seven different time snapshots. Taking the mean of all values of $\omega_2$ from $R=4$ to $R=15$Kpc in the stack plot, we define a pattern speed $\Omega=21$Km s$^{-1}$Kpc$^{-1}$ (outer red line, superposed to all panels of Fig.\ref{fig:meanome2}). However, at radii $R<4$Kpc, a second pattern speed is also identifiable, although its corresponding secondary plateau in the curve $\omega_2(R,t)$ exhibits noisy fluctuations and it is not as conspicuous as in the case of the main plateau. By the same procedure, we recognize a second pattern speed $\Omega_p^{(2)}=37$Km s$^{-1}$Kpc$^{-1}$. This represents a central bar-like oval distortion which rotates faster than the outer spirals. 

Similar features are observed in the $m=2$ patterns formed in the time interval of Fig.\ref{fig:meanome1}. In this case, the outer plateau of $\omega_2$ appears less clear and less extended, appearing, in particular, to form a break at a distance of $\approx 10$Kpc, and presenting also some time fluctuation of its level value. Also, at distances below the inner breaking point, which in this case too is near $4$Kpc, the existence of a well-defined second pattern speed is less clear as well. Examining several experiments, we find that, in general, the various pattern speeds defined by the above method are always more clear when the amplitude of the $m=2$ mode is also larger. This is consistent with the nature of a density wave, i.e., the wave tends to acquire a quasi-stationary form as its self-gravity increases. At any rate, in both Figs.\ref{fig:meanome1} and \ref{fig:meanome2}, the innermost part of the spirals (roughly from 4 to 9 Kpc) appears to move with a well defined, constant with the distance $R$, pattern speed, which, furthermore, appears to have quite similar value in both time intervals. Thus, our spirals in this domain of the disc tend to acquire a `quasi-stationary' form within the considered time intervals. 
\begin{figure}
\begin{center}
\includegraphics[scale=0.42,angle=-90]{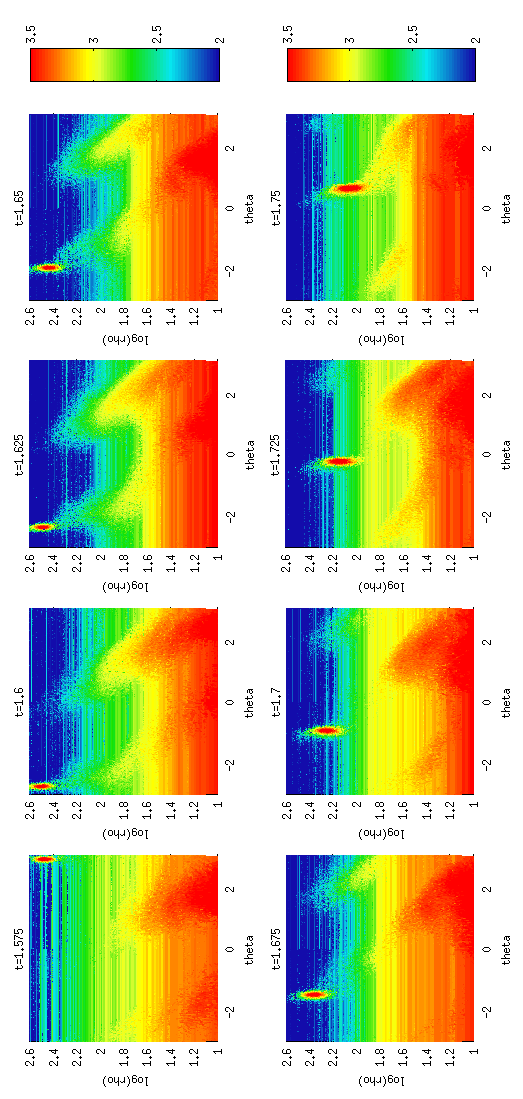}
\end{center}
\caption{A color map of the projected surface density of the disc $\Sigma_d(R,\phi)$ in the co-ordinate representation $(\phi,\ln R)$ at eight different time snapshots. The features extending outwards with nearly constant slope indicate logarithmic spiral arms. The small ellipse in the upper part of each plot is the moving satellite.}
\label{fig:logplot}
\end{figure}

\begin{figure}
\begin{center}
\includegraphics[scale=0.42,angle=0]{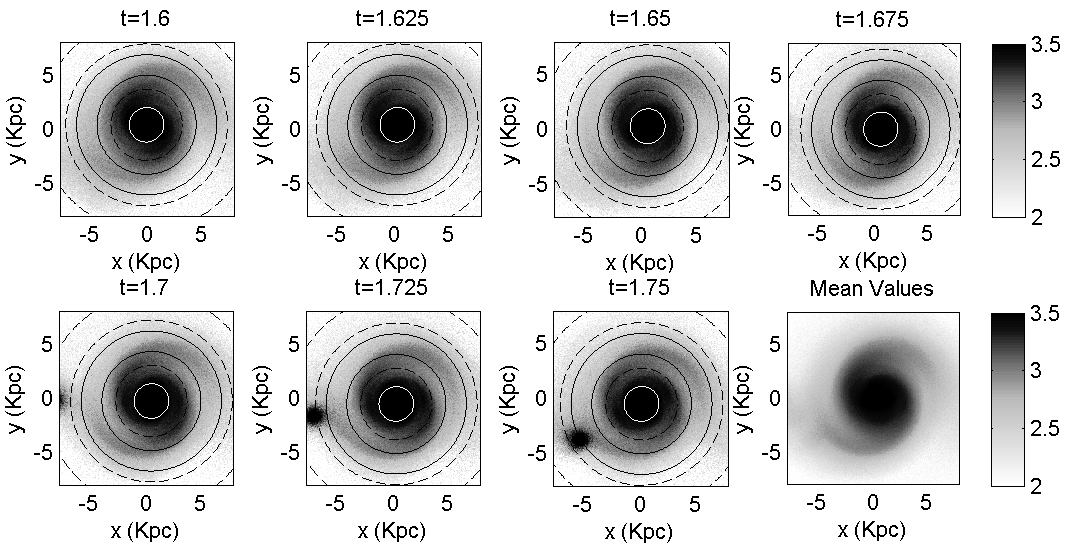}
\end{center}
\caption{Seven consecutive time snapshots of the disc for the times $t=1.6 + i\Delta t$, with $\Delta t=0.025$Gyr, each snapshot having been rotated clockwise with respect to the true orientation by an angle equal to $i\Omega_p \Delta t$, $i=0,\ldots,6$, with $\Omega_p$ as specified by the Fourier analysis of the spiral mode. The last panel is a stack plot of all seven previous plots. The three dashed circles indicate the positions of the Inner Lindblad Resonance, corotation, and the Outer Linblad Resonance, specified with the spiral pattern speed. The red circles indicate corotation and the Outer Lindblad resonance as specified by the inner bar pattern speed.}
\label{fig:stackplot}
\end{figure}

More features of the observed spiral arms can be unraveled with the help of Figs.\ref{fig:logplot} and \ref{fig:stackplot}. The relevant information is contained in the projected surface density maps $\Sigma_d(R,\phi)$, viewed first, in Fig.\ref{fig:logplot}, in a $(\phi,\log(R))$ co-ordinate representation. These plots indicate that: i) the spiral pattern speed is different from the local angular speed of the satellite. This fact notwithstanding, tidal tails moving with a range of intermediate values of the angular speed connect the outer parts of the spiral arms with the satellite. ii) The spiral arms are nearly logarithmic, with an estimated pitch angle $i_0=\tan^{-1}d\ln R/d\phi_2\approx 21^\circ$ in the panels of Fig.\ref{fig:logplot}. 

Having determined the spiral pattern speed as $\Omega_p=21$Kms$^{-1}$Kpc$^{-1}$, Fig.\ref{fig:stackplot} shows seven successive different snapshots of the main galaxy's disc, for the times $t=1.6+i\Delta t$, $i=0,1,\ldots,6$, each rotated around the disc central axis by the angle $\Delta\phi = - i\Omega_p\Delta t$, with $\Delta t=0.025$. We note immediately the quasi-invariance (besides rotation) of the spiral pattern in a range of distances $4<R<9$Kpc; the pattern is preserved in every one of the depicted individual snapshots as well as in their superposition as a stack plot. In comparison, the inner bar-like oval distortion still appears to rotate from one snapshot to the other, with a relative (with respect to the spiral) pattern speed equal to $\Omega_p^{(2)}-\Omega_p\simeq 16Kms^{-1}Kpc^{-1}$. The near invariance of the spiral arms in the domain $4<R<9$ Kpc, as well as their nearly constant pattern speed, is an indication that these arms are density waves. The near invariance of their shape is best unraveled by creating a stack plot of all seven snapshots (last panel in Fig.\ref{fig:stackplot}). This smears out the central bar, which appears now like a bulge. However, the spiral pattern seen in the stack plot is invariant with respect to the one seen in any individual time snapshot in the same domain of the disc. 

Finally, in the same figure we give the approximate position of disc resonances in comparison to the extent of the bar and spiral modes in the indicated time interval. Using the potential data on the N-body grid, we compute by interpolation an azimuthally averaged gravitational potential $V_0(R)$ as a function of $R$ accross the disc. This allows, in turn, to compute the azimuthal and epicyclic frequencies $\Omega(R)$ and $\kappa(R)$ (Eqs.(\ref{eq:epic})), and hence the positions of the inner and outer Lindblad resonances, as well as corotation, i.e., the values of $R$ where:
\begin{equation}\label{iolr}
\Omega(R)-{1\over 2}\kappa(R)=\Omega_p~~\mbox(ILR),~~~
\Omega(R)=\Omega_p~~\mbox(CR),~~~
\Omega(R)+{1\over 2}\kappa(R)=\Omega_p~~\mbox(OLR)
\end{equation}
The three dashed circles in each panel of Fig.\ref{fig:stackplot} show the position of the ILR, corotation, and OLR (from inside to outside) when the pattern speed $\Omega_p$ is assigned the spiral value $21$Km s$^{-1}$Kpc$^{-1}$, while the two red circles indicate corotation and the OLR when, instead, we use the second pattern speed value referring to the bar, i.e., $\Omega_p=\Omega_p^{(2)}=37$Km s$^{-1}$Kpc$^{-1}$. We see that the spiral mode appears to start close to the ILR (computed with the spiral pattern speed $\Omega_p$), while it ends a little inside the corotation radius of the same pattern. This is consistent with the extent of spiral arms in so-called `precessing ellipses' models of spiral structure (see e.g. \citet{Contopoulos:1, Patsis:2}). On the other hand, we did not find any evidence of coupling between the bar resonances and the spiral ones, or, more generally, the bar resonances and the morphological features of the observed spiral arms. 

\section{Conclusions}
We used a newly-developed N-Body code (MAIN) in order to study, by numerical simulations, the features of non-axisymmetric structures induced in a galactic disc via the mechanism of repeated flybys of a companion satellite galaxy. We made a suite of simulations, varying the disc's temperature (Q-profile) and the mass and initial distance (major semi-axis) of the satellite, chosen to be in an orbit of relatively low initial inclination ($30^\circ$), and initial eccentricity $e=0.4$. The satellite orbitally decays due to dynamical friction caused by its interaction with the live halo of the main galaxy. Our main focus was on characterizing the space and time features of the spiral structure induced on the disc due to the mechanism of repeated (typically 4 to 5, before merger) flybys of the companion. A summary of our methods and results in the previous sections is the following: 

1. On purely numerical ground, we exploited the fast convergence properties of a new Cartesian mesh-type solver of Poisson's equation in three-variables configuration space, which is the core of the MAIN code. The fast convergence and near optimal performance is due to the use of the new Symmetric Factored Approximate Sparse Inverse (SFASI, documented in \citet{Kyziropoulos:1}) preconditioning scheme in conjunction with the multigrid method. Boundary conditions are computed via multipole expansions of the potential, which have a computational complexity linear in the number of particles, while the Poisson solver has a computational complexity close to $O(n)$ where $n$ is the number of grid points (see section 2).

2. We examined the role of two key parameters of the `repeated flyby' mechanism, namely, the initial satellite to disc mass ratio, and the temperature (asymptotic Q-value) of the disc, in the mechanism's efficiency with respect to a particular phenomenon, i.e., the generation of bi-symmetric spiral arms in the disc. Regarding the initial satellite to disc mass ratio, we found that, for our specific initial conditions, the repeated flyby mechanism becomes efficient in a rather narrow band of values (close to 1:5 for our specific initial conditions). For lower rations, the mechanism appears rather ineffective, while, for larger ones, it is effective but only short-lived: the satellite's orbital decay towards the disc becomes quite rapid. Regarding the disc's Q-value, we noted the importance of the antagonism between self-induced and companion-induced modes on the disc. Again, in our specific initial conditions, the self-induced modes grow rapidly enough to prevail over the companion-induced modes when $Q=1.5$, while the inverse phenomenon takes place when $Q=2$. On the other hand, the disc becomes all together non-responsive to either type of excitations when $Q$ is set as high as $Q=2.5$.

3. Finally, we gave some features of the form and evolution of the companion-induced spiral structures in those experiments in which the latter are observed to be relatively long-lived, i.e., persisting for 4 to 5 revolutions. We found that our companion-induced spiral modes are well approximated with logarithmic spirals keeping a nearly constant pattern speed in a disc's domain extending from about the Inner Lindblad Resonance to near, but inside, corotation. There are also companion induced bar modes inside the ILR, which typically rotate faster than the spirals. The spirals in the above domain have features of a self-gravitating density wave, while spiral extensions beyond corotation may also be observed joining the satellite via tidal bridges. 

\section*{Acknowledgements}
This research was supported in part by the Academy of Athens Research Committee, project 200/854.

\appendix
\section{Poisson Solver based on the Multigrid method in conjuction with the SFASI matrix}
\begin{figure}
	\begin{center}
	\includegraphics[scale=0.5]{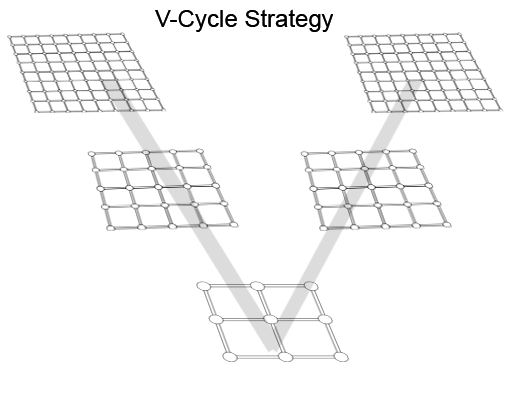}
	\end{center}
	\caption{The Vcycle Multigrid strategy}
	\label{fig:figure2}
\end{figure}

The Multigrid method has used extensively in Computational Physics due to its near-optimal complexity and its convergence behavior, \citep{a:6, a:10, a:5, b:3, b:5, a:3, a:7, b:7}. The Multigrid method exploits the way stationary iterative schemes handle the error by creating a hierarchy of (finer and coarser) grids, transferring the linear system to various levels with different mesh sizes, i.e:
\begin{equation}
	A_{\ell}x_{\ell} =b_{\ell}
\end{equation}
where ${\ell}$ denotes the hierarchy grid levels. The method is composed of three distinct components:\\
\textbf{1. Transfer Operators:} These are special operators that are used to transfer vectors from coarser (2h) to finer (h) grids (Prolongation) and finer (h) to coarser grids (2h) (Restriction); for a tutorial and introduction to the multigrid method see \citet{b:3, b:5}. The full-weighting operator used as Restriction is given in stencil notation by,
\begin{equation}
R=\frac{1}{64}\begin{bmatrix}
1 & 2 & 1\\ 
2 & 4 & 2\\ 
1 & 2 & 1
\end{bmatrix}^{2h}_{h}\begin{bmatrix}
2 & 4 & 2\\ 
4 & 8 & 4\\ 
2 & 4 & 2
\end{bmatrix}^{2h}_{h}\begin{bmatrix}
1 & 2 & 1\\ 
2 & 4 & 2\\ 
1 & 2 & 1
\end{bmatrix}^{2h}_{h}
\end{equation}
and the Tri-linear interpolation used as Prolongation can be expressed in stencil notation, \citep{b:3, b:5}, by,\\
\begin{equation}
P=\frac{1}{16}\left ]
\left ]\begin{matrix}
1 & 2 & 1\\ 
2 & 4 & 2\\ 
1 & 2 & 1
\end{matrix}\right [^{h}_{2h}
\left ]\begin{matrix}
2 & 4 & 2\\ 
4 & 8 & 4\\ 
2 & 4 & 2
\end{matrix}\right [^{h}_{2h}
\left ]\begin{matrix}
1 & 2 & 1\\ 
2 & 4 & 2\\ 
1 & 2 & 1
\end{matrix}\right [^{h}_{2h}
\right [
\end{equation}
A sparse matrix representation of these operators is used and thus the transfer procedure between the levels results to matrix by vector multiplication.\\
\textbf{2. Cycle Strategy:} The cycle strategy as depicted in Figure \ref{fig:figure2}, refers to the sequence in which the grids are visited and the respective corrections are obtained, \citep{a:10, b:3, b:5}. Here the V cycle strategy is used, where the method descends to coarser levels executing   pre-smoothing iterations at each level and then the method ascends executing   post-smoothing iterations at each level.\\
\textbf{3. Smoother:} Let us consider the Preconditioned Richardson's method, \citep{a:10, b:3, b:5}:
\begin{equation}
	x_{\ell}^{(k+1)}=x_{\ell}^{(k)}+\omega M_{\ell}(b_{\ell}-A_{\ell}x_{\ell}^{(k)}), k=0,1,2...
\end{equation}
where
\begin{equation}
	A_{\ell}=U_{\ell}^TU_{\ell}+E
\end{equation}
The $M_{\ell}$ matrix denotes the SFASI approximate inverse preconditioner and ${\omega}$ is the damping parameter ${0< \omega <2}$. The Symmetric Factored Approximate Sparse Inverse matrix, namely (SFASI), \citep{Kyziropoulos:1}, is based on the incomplete Cholesky-type factorization, \citep{a:14, a:13}. Moreover, approximate inverse sparsity patterns, \citep{a:15, a:16}, are used in conjunction with a restricted solution process minimizing the search for elements when computing the factor of the approximate inverse. The SFASI approximate inverse preconditioner is given by:
\begin{equation}
	M_{\ell}=((G^{lfill}_{droptol})(G^{lfill}_{droptol})^T)_{\ell}
\end{equation}
where G  is the upper triangular factor of the SFASI approximate inverse preconditioner, lfill denotes the levels of fill (i.e. the number of levels of neighboring vertices kept for each vertex of the graph of matrix U) and drptol the predetermined drop tolerance used to sparsify the upper factor U corresponding to the coefficient matrix A.
Furthermore, the Dynamic Over/Under Relaxation (DOUR) algorithm is used to determine the smoothing parameter. Given an initial damping parameter ${\omega}$, at (k+1) iteration, the solution is given by, \citep{c:2}:
\begin{equation} \label{eq:2}
x_{\ell}^{(k+1)}=x_{\ell}^{(k)}+\kappa\Delta x_{\ell}^{(k)},
\end{equation}
where
\begin{equation} \label{eq:3}
\Delta x_{\ell}^{(k)}=\tilde{x}_{\ell}^{(k)}-x_{\ell}^{(k)},
\end{equation}
and
\begin{equation} \label{eq:4}
\kappa =\frac{(\Delta x_{\ell}^{(k)})^T(b_{\ell}-A_{\ell}\tilde{x}_{\ell}^{(k)})}{(\Delta x_{\ell}^{(k)})^TA_{\ell}\Delta x_{\ell}^{(k)}}.
\end{equation} 
From (\ref{eq:2}), (\ref{eq:3}) and (\ref{eq:4}), we obtain
\begin{equation} \label{eq:5}
x_{\ell}^{(k+1)}=x_{\ell}^{(k)}+\omega_{e}((G^{lfill}_{droptol})(G^{lfill}_{droptol})^T)_{\ell}(b_{\ell}-A_{\ell}x_{\ell}^{(k)})
\end{equation}
where ${\omega_{e}=\omega(1+\kappa)}$ is the effective relaxation parameter and equation (\ref{eq:5}) is the proposed two stage non-stationary smoothing scheme. Further information and convergence analysis of the DOUR algorithm are given in \citet{c:2}.\\
The V-Cycle multigrid algorithm in conjunction with the SFASI matrix as smoother is given by the following algorithmic compact scheme:
\newcommand{\tab}[1]{\hspace{.05\textwidth}\rlap{#1}}
\begin{itemize}
\item ${x_{\ell}\rightarrow \textbf{MGV}(A_{\ell},(G^{lfill}_{droptol})_{\ell},x_{\ell},b_{\ell},\ell)}$\\
\item \textbf{IF} ${{\ell}={\ell}_c}$ is the coarsest level\\
\item \tab{${x_{{\ell}_c}^{(k+1)}=x_{{\ell}_c}^{(k)}+\omega (G^{lfill}_{droptol})_{{\ell}_c}(G^{lfill}_{droptol})^{T}_{{{\ell}_c}}(b_{{\ell}_c}-A_{{\ell}_c}x_{{\ell}_c}^{(k)}),~~~k=1,2,...,v_3}$}\\
\item \textbf{ELSE}\\
\item \tab{${x_{{\ell}}^{(k+1)}=x_{{\ell}}^{(k)}+\omega (G^{lfill}_{droptol})_{{\ell}}(G^{lfill}_{droptol})^{T}_{{\ell}}(b_{{\ell}}-A_{{\ell}}x_{{\ell}}^{(k)}),~~~k=1,2,...,v_1}$}\\
\item \tab{${b_{{\ell-1}}\leftarrow R(b_{\ell}-A_{\ell}x_{\ell})}$}\\
\item \tab{${x_{{\ell-1}}\leftarrow 0}$}\\
\item \tab{${x_{{\ell-1}}\rightarrow \textbf{MGV}(A_{{\ell-1}},(G^{lfill}_{droptol})_{{\ell-1}},x_{{\ell-1}},b_{{\ell-1}},\ell-1)}$}\\
\item \tab{${x_{{\ell}}\leftarrow x_{\ell}+P(x_{{\ell-1}})}$}\\
\item \tab{${x_{{\ell}}^{(k+1)}=x_{{\ell}}^{(k)}+\omega (G^{lfill}_{droptol})_{{\ell}}(G^{lfill}_{droptol})^{T}_{{\ell}}(b_{{\ell}}-A_{{\ell}}x_{{\ell}}^{(k)}),~~~k=1,2,...,v_2}$}\\
\item \textbf{END}
\end{itemize}

\bibliographystyle{mn2e}
\bibliography{references}

\bsp	
\label{lastpage}

\end{document}